\documentclass[manuscript]{emulateapj}
\def\h2{H$_2$\space}

\def\micron{$\mu$m}
\def\microns{$\mu$m\space}
\def\arcsec{$^{\prime\prime}$}
\def\arcsecs{$^{\prime\prime}$\space}
\def\arcmins{$^{\prime}$\space}
\def\arcmin{$^{\prime}$}
\def\deg{$^{\circ}$}
\def\degs{$^{\circ}$\space}

\def\Msun{M$_\odot$\space}
\shorttitle{  {\it Spitzer}  images of Cep E}
\shortauthors{Velusamy et al.}

\begin{document}

\title{Jets and Wide Angle Outflows in Cepheus E: New Evidence from {\it Spitzer} }

\author{T. Velusamy\altaffilmark{1}, W. D. Langer\altaffilmark{1}, M. S. N. Kumar\altaffilmark{2},  J. M. C. Grave\altaffilmark{2} }

\altaffiltext{1}{Jet Propulsion Laboratory, California Institute of
Technology, 4800 Oak Grove Drive, Pasadena, CA 91109;
velusamy@jpl.nasa.gov, William.D.Langer@jpl.nasa.gov }
 \altaffiltext{2}{Centro de Astrofisica da Universidade do Porto, Rua das Estrelas, 4150-762 Porto, Portugal; nanda@astro.up.pt, jgrave@astro.up.pt}


\begin{abstract}
 Outflows and jets are believed to play a crucial role in determining the mass of the central protostar and its planet forming disk by
virtue of their ability to transport energy, mass, and momentum of the surrounding material, and thus terminate the infall stage in star
and disk formation. In some protostellar objects both wide angle outflows and collimated jets are seen, while in others only one is
observed. {\it Spitzer} provides unprecedented sensitivity in the infrared to
study both the   jet and outflow features.  Here, we use    HiRes deconvolution to improve the
visualization of spatial morphology by enhancing resolution (to
sub-arcsecond levels in the IRAC bands) and removing the
contaminating sidelobes from bright sources. We apply this approach to study the jet and outflow features in Cep E a young, energetic Class 0 protostar.  In the reprocessed images we detect: (i) wide angle outflow seen in scattered light; (ii) morphological details on at least 29 jet driven bow shocks and jet heads or knots; (iii) three compact features in 24 \microns continuum image as atomic/ionic line emission  coincident with the  jet heads; and, (iv) a flattened $\sim$35\arcsecs size protostellar envelope seen  against the interstellar background PAH emission  as an absorption band  across the protostar at 8 \micron.  By separating the protostellar photospheric scattered emission in the wide angle cavity from the jet emission we show that we can study directly the scattered light spectrum.  We present the \h2 emission line spectra, as observed in all IRAC bands,   for 29 knots in the jets and bowshocks and use them in the IRAC color -- color space  as a diagnostic of the thermal gas in the  shocks driven by the jets. The data presented here will enable detailed modeling of the individual shocks   retracing the history of the episodic jet activity and the associated accretion on to the protostar.  The {\it Spitzer} data analysis presented here   shows the richness of its archive as a resource  to study    the jet/outflow features in \h2 and scattered light in a large homogeneous sample.

\end{abstract}

\pagebreak
\section{INTRODUCTION}
Star formation begins with the collapse of a dense interstellar cloud core to a protostar surrounded by a disk of gas and dust.  The envelope of the cloud core falls inwards to feed further growth of the protostar and its accreting disk.  At some point during the accretion phase, protostellar jets and winds originate close to the surface of the forming star and interact with the dense envelope surrounding the protostar-disk system (e.g.  K\"{o}nigl \& Pudritz  2000; Shu et al. 2000). These outflows and jets are believed to play a crucial role in determining the mass of the central protostar and its planet forming disk by
virtue of their ability to transport energy, mass, and momentum of the surrounding material, and thus terminate the infall stage in star
and disk formation. It is now well recognized that jets and outflows are essential and inherent to  the star formation process.  These flows  are broadly classified into two types: molecular outflows traced mainly with CO emission, and jets observed by optical emission (cf. recent reviews by Arce et al. 2007, Pudritz et al. 2007; Bally 2007).  Observations indicate that around some protostars, high-velocity jets with a narrow opening angle are enclosed by a low-velocity outflow with a wide opening angle (e.g.  Mundt \& Fried 1983;  Velusamy et al. 2007) however, in others only one component is observed. Poorly collimated flows can be due to extreme precession of the jet (Shepherd et al. 2000) and indistinguishable from wide angle outflows.  The wide angle outflows, which are also observed in the scattered light, are often missed in the optical or near-IR due to instument insensitivity to faint emission   and confusion from the bright protostellar emissions.   {\it Spitzer} provides unprecedented sensitivity in the infrared to
detect the jet/outflow features.    The IRAC, IRS, and MIPS observations can provide new insight into the structure, morphology, and physical and chemical characteristics of the outflow sources as demonstrated in the observations of HH46/47  by Noriega-Crespo et al. (2004a) \& Velusamy et al. (2007),   Cep E by Noriega-Crespo et al. (2004b),  L1512F by Bourke et al. (2006), L1448 by Tobin et al. (2007) \& Dionatos et al. (2009),    L1251 by Lee et al. (2010), and  HH211 by Dionatos et al. (2010).  Using the IRAC colors alone, Ybarra \& Lada (2009) demonstrated the feasibility to study the pure rotational \h2 line emissions in the shocks without the need for spectroscopic data.  Velusamy et al. (2007) demonstrated that resolution enhanced re-processing of {\it Spitzer} images in the case of  HH46/47  brought  out more clearly the morphologies of: (a)  wide-angle outflow cavities in the IRAC 3.6 \& 4.5 \microns images by scattered starlight, (b) jets and bow shocks in \h2 molecular emission within the IRAC bands, and (c) the hottest atomic/ionic gas in the jet head  characterized by [Fe II]  and [S I] line emissions within the MIPS 24 \microns band.

In this paper we present new results on  Cep E  obtained by reprocessing the {\it Spitzer} IRAC and MIPS images  with the  HiRes deconvolution (Backus  et al. 2005).
Cep E is a remarkably bright outflow object, at a distance of 730 pc, with strong radiative power ($\sim$ 80 L$_\odot$), and appears to be an intermediate mass object which is expected to reach a final stellar mass of 3M$_\odot$ (Froebrich  et al. 2003). Cep E has been recently classified as a Class 0 protostar  by Noriega-Crespo et al. (2004b) based on the {\it Spitzer} Early Release Observations. It is among the youngest protostars with an  age   in the range of 20 -- 400 thousand years, estimated for different evolutionary models (Froebrich, 2005; and the references therein). In the near- and mid-IR  its morphology is similar to that expected for a jet driven outflow, where the leading bow shocks entrain and accelerate the surrounding molecular gas.  The properties of this  outflow have been well studied by Moro-Mart\'{i}n et al. (2001) and Smith et al. (2003).  The southern bright bowshock, identified as HH 377 (Devine, Reiputh \& Bally 1997), is well studied in the optical wavelengths using the  [O I] $\lambda$ 6300, [NI ] $\lambda$ 5200 and [S II] $\lambda\lambda$6717,6731 lines tracing the dissocitive shocks characteristics of HH flows (e.g. Ayala et al. 2000) and later in the near-IR in the [Fe II] 1.26 and 1.64 \microns emission lines, utilizing the fact that  [S II] and [Fe II] have very similar excitation energies  (Nisini et al. 2002). Furthermore,  Hartigan, Raymond \& Hartmann 1987) interpreted the relative strength of [O I] and [N I] with respect to H$_\alpha$ and the   lack of [O II] and [O III] in terms of    relatively low velocity (20-25 km s$^{-1}$) dissociative shocks and concluded that  the south bowshock of Cep E  is a low excitation HH object.    In addition to dissociative shocks, the   HH outflows are known to have also magnetic precursors (C-shocks) needed to explain excitation of the molecular
H$_2$ emission. Cep E lobes have been well imaged in H$_2$ emission in IR: in v=1--0 S(1)  2.122\micron; v=2--1 S(1) 2.248 \micron; v=3--2 S(3) 2.202 \microns (Eisl\"{o}ffel et al. 1996; Ladd \& Hodapp, 1997); v=0--0 S(5) 6.91 \microns and S(3) 9.66 \microns (Noriega-Crespo 1998); v=1--0 S(1)1.21 \micron. In the mid-IR a series of 6 pure rotational emission lines H$_2$ 0-0  S(2) to S(7) in the wavelength range of 12.28 \microns to 5.51 \microns
 in the Cep E lobes were first imaged by  ISOCAM CVF (Moro-Mart\'{i}n et al. 2001) and their excitation diagrams led to the use of H$_2$ line ratios as C-shock diagnostics.  The emissions in the Spitzer images in all the IRAC bands are dominated by a wide range of H$_2$ rotational transitions from 0--0 S(4) to 0--0 S(13) and a few vibrational 1--0 transitions in the 3.6 \microns band. Thus Spitzer observations of Cep E enhance  the rich data set already  available in optical, near-IR and mm wavelengths. Furthermore,
   all the interpretations so far have focused on treating this object as having jets and bow shocks, but no wide angle outflow.  Because of the limited spatial resolution even the CO outflow morphology has been interpreted as entrained low velocity molecular gas associated with the bow shock driven by the jets (Moro-Mart\'{i}n et al. 2001).  Our enhanced image analysis   show  the simultaneous presence of wide angle outflow along with the bow shocks powered by the jets.  In this paper  we highlight  the morphological features in the low velocity wide angle outflow and the high velocity jets  and discuss a wide range of structures from compact knots to extended envelopes as seen in the reprocessed {\it Spitzer} IRAC and MIPS images.

\section{ANALYSIS}

  Noriega-Crespo et al. (2004b) presented    the  mosaic images of Cep E in the {\it Spitzer} IRAC bands at 3.5 \micron, 4.5 \microns and 8 \microns in a three color representation  and the mosaic image of MIPS band at 24 \micron. However, some of the structures were missed in these mosaic maps  due to either lower spatial resolution or diffraction lobes.
   {\it Spitzer} provides unprecedented sensitivity in the infrared, but the spatial resolution is limited by the relatively small aperture (0.85 m) of the primary mirror.  To maximize the scientific return of {\it Spitzer} images we  use the HiRes deconvolution processing technique that makes the optimal use of the spatial information in the observations.    The algorithm, ``HiRes'' and its implementation has been discussed by Backus et al. (2005) and its performance on a variety of astrophysical sources observed by {\it Spitzer} is presented by Velusamy et al. (2008).  The HiRes deconvolution algorithm is based on the Richardson-Lucy algorithm (Richardson 1972; Lucy 1974), and the Maximum Correlation Method employed by Aumann et al. (1990) for IRAS data.  As demonstrated by Velusamy et al. (2007; 2008)  the HiRes deconvolution on {\it Spitzer} images retains a high fidelity preserving all the main features.  HiRes deconvolution improves the visualization of spatial morphology by enhancing resolution and removing the contaminating sidelobes from bright sources.  The benefits of HiRes include: (a) enhanced resolution  $\sim$ 0.6\arcsecs -- 0.8\arcsecs for IRAC bands; $\sim$ 1.8\arcsecs and $\sim$7\arcsecs for MIPS 24 \microns and 70 \micron, respectively;    (b) the ability to detect sources below the diffraction-limited confusion level; (c) the ability to separate blended sources, and thereby provide guidance to point-source extraction procedures; (d) an improved ability to show the spatial morphology of resolved sources.
   We reprocessed the images at all the IRAC bands and the MIPS 24 and 70\microns bands using all available map data in these bands in the {\it Spitzer} archives containing the prostellar object Cep E centered at ($\alpha$ = 23h03m12.5s, $\delta$ = +61\deg42\arcmin30\arcsec [J2000]). We used the pipeline processed basic calibrated data (BCD)  down-loaded from the {\it Spitzer} Science Center (SSC) archives. We constructed    mosaic images at each band using the MOPEX  software (provided by SSC). In addition, we also produced reprocessed images at each band by applying the HiRes deconvolution  on BCDs following the steps outlined by Velusamy et al. (2008). The mosaic and the HiRes images  cover  an angular size of 5\arcmins centered on the Cep E protostar.   All the HiRes images shown in this paper were obtained after 50 iterations unless mentioned otherwise.

The mosaic and the HiRes deconvolved images in the  IRAC and MIPS bands, shown in Figures 1 and 2 respectively, bring out the merits of HiRes processing. In these figures the top panels show the reprocessed HiRes images and the corresponding observed (mosaic) images are shown in the lower panels.  The mosaic and HiRes images have identical angular extent and scale.  The images are presented  using a square-root intensity scale and color stretches. The color stretches used
are similar in all images and an example of the   colors representing  the relative
intensities is shown by the wedge at the upper left.
In the MIPS 24 \microns mosaic image only the protostar is obvious.
However, in the HiRes image at least three additional features along the atomic jet become identifiable  as   sources distinct from the ``side lobes'' present in the mosaic image.  These were not identified   in   previously published mosaic images.

The MIPS images shown in Fig. 2 have small pointing differences ($\sim$ 1\arcsec) with respect to the IRAC images in Fig. 1.  A comparison of the point sources in the MIPS 24 \microns image with their counterparts in the IRAC bands indicated that the  pointing at 24 \microns was off with respect to the IRAC bands by -0.07$^s$ in RA and +1.1\arcsecs in Declination. We shifted the  MIPS 70 \microns map position by matching the protostar position with the MIPS 24 \microns image.   In all subsequent displays and analysis we use the MIPS images corrected for their position shifts with respect to the IRAC images.

\section{RESULTS AND DISCUSSION}
The {\it Spitzer} IRAC and MIPS bands offer a unique
resource to study a wide range of components simultaneously: protostars, protostellar disks, outflows,
protostellar envelopes and cores. At the short wavelength IRAC bands the
outflow cones are observable in scattered light from the protostar
through the cavity created by the jets and outflows (cf. Tobin et al.
2007; Velusamy et al. 2007). A significant fraction of the emission in the IRAC bands is
also considered to contain emission
 from the  H$_2$ rotational lines
 (e.g. Noriega-Crespo et al. 2004a,b;  Neufeld \& Yuan, 2008; Smith \& Rosen, 2005) and therefore is an excellent tracer of  \h2 emission in the  protostellar jets. Recently Velusamy et al. (2007)
  have shown that  molecular jets and molecular gas in the bow shocks are readily identifiable
  in the IRAC bands, while the hottest atomic/ionic gases in
the bow shocks, are also identifiable  in the MIPS 24 \microns band which covers a few
atomic/ionic  emission lines. The dust emission from the protostar in
the MIPS bands is a good diagnostic of the circumstellar disks.  Finally, an extended   flattened infall envelope perpendicular to the outflow has been seen in the IRAC 8 \microns  in absorption against the background interstellar PAH emission (Looney et al. 2007; Tobin et al. 2010). {\it Spitzer}'s  coverage of a broad range of   IR emissions in the IRAC and MIPS bands with high sensitivity and  photometric stability along with the IRS data in some cases,  provide sufficient data for a comprehensive modeling of the SED    to derive the physical characteristics and  the evolutionary stages of protostars (cf. Robitaille et al. 2007; Forbrich et al. 2010). However, by applying the HiRes re-processing we can extract even more information from the {\it Spitzer} data. The sensitivity, the resolution enhancement and removal of the diffraction lobe confusion have led to visualizing and characterizing more clearly the following features in the  Cep E protostar:
\begin{enumerate}
\item very high dynamic range maps in which  the protostar is well resolved on a sub-arcsec scale from the surrounding jets and outflows
\item wide-angle outflow cavities in the IRAC 3.6 and 4.5 \microns images identified by the scattered photospheric starlight; Though they are visible in the Mosaic (standard PostBCD products), they are brought out more clearly in the HiRes images.
\item jets and bow shocks in \h2 molecular emission within the IRAC bands,
\item hottest atomic/ionic gas in the jet head traced by the [Fe II]/[S I] line emissions within MIPS 24 \microns continuum band.
\item flattened infall envelope around the protostar seen in extinction at 8 \micron.
\end{enumerate}

\subsection{Protostar}
The protostar in Cep E is detected in all IRAC and MIPS bands.  The IRAC  HiRes images  shown in Fig. 1 have  high dynamic range, but the brightest features are shown saturated as they are optimized to highlight the low surface brightness emission.  In all bands the highest brightness is  at the protostar location while the jets and the outflow features are at very low brightness, at  levels of $\sim$ 1\% or less of the peak brightness.  The color stretch  used in Fig. 1 is designed to bring out these low surface brightness features and therefore, underrepresents the peak brightness at the protostar.  In Figure 3 we show the HiRes intensity contour maps of the protostar in all bands. To bring out the huge intensity contrast between the brightest emission at the  protostar and the weaker emission of the  outflow-jet features plotted in Fig. 3, we show selected contours: the lowest  level at 0.01\% and the highest at 50\% of the peak.  The observed flux densities and sizes  for the protostar are summarized in Table 1. In the IRAC wavelengths the emission from the protostar appears to be  unresolved with sizes   comparable to  the  HiRes beam.   The protostar flux densities measured in the HiRes maps listed in Table 1 are consistent with the photometric estimates discussed below in section 3.4.

 Cep E is believed to be a double source separated by 1.4\arcsecs ($\sim$1000AU) as shown by the 222 GHz millimeter interferometric observations of Moro-Mart\'{i}n et al. (2001).  However, as seen in Fig. 3, the overlay of mm source positions on the IRAC 3.6 \microns map  does not show any evidence for a double source in the IR bands.  The sub-arcsec resolution ($\sim$ 0.6\arcsecs -- 0.8\arcsec) in the HiRes images would have clearly resolved  any double source if present. Though the protostar emission is barely resolved in the maps, there is a hint of slight  elongation roughly along the separation between the mm-double sources.  Our results do not show any evidence for the presence of  a second protostellar component. Instead, we suggest that the resolved  mm double source structure may represent circumstellar cold dust within a few $\times$10$^3$ AU.

\subsection{Wide Angle Outflows and Jets: IRAC Images}

It can be difficult to observe the extended low surface brightness features associated with the outflow or jets    as they may be severely
confused by the high brightness of the protostar and the disk.
Deconvolution at best removes, or at least  minimizes, the diffraction
effects of the brightest features, thus  enabling improved visualization of
the low surface features around them. Furthermore the resolution
enhancement provides a sharper view of the outflow cavity walls,
 molecular jets, and bow shocks.
\subsubsection{Wide Angle Outflow Component}
   The radiation from deeply embedded young protostars is completely opaque at short wavelengths and we can observe only the scattered light escaping along the bipolar cavities produced by their jets and outflows.   The scattering models (cf. Whitney et al. 2003a \& b; Tobin et al. 2007) show that such outflow cavities can be traced in {\it Spitzer} images in IRAC  3.6 and 4.5 \microns bands, as is seen more clearly in the HiRes deconvolved images (Velusamy et al. 2007; 2008).  In Fig. 1 a wide-angle biconical outflow is clearly detected in the HiRes images at 3.6 and 4.5 \micron.  Though its presence  is marginally evident in the  mosaic images, its morphological details are  delineated more clearly  only in the HiRes images.  The wide angle outflow is clearly detected in the light from the protostar scattered by the outflow cavity.  Scattered light is brightest at the 3.6 and 4.5 \microns bands. As expected for the photospheric light emitted by  a protostar at T $\sim$ 4000K (see section 3.4), the scattered light tracing the wide angle outflow component is relatively weak  at 5.8 \microns and nearly absent in the  8 \microns image.  In contrast to this wide angle outflow traced by the scattered light from the protostar, the narrower and collimated jets  and the associated bow shocks are detected by their \h2 emission in all the IRAC bands.  All the four IRAC bands contain a few \h2 spectral lines (see below for details). While bands 1 and 2 show both the wide angle outflow (in scattered light) and narrow jet features (in \h2 emission),  the bands 3 and 4 are dominated by the latter. In the HiRes images the jet emission features show well resolved knots and bow shocks. In the north lobe alone there are at least 16 such features (see Fig. 5a).

The HiRes processing brings out the morphology of the jet and outflow more clearly.  In Fig. 4 we show a comparison between  the 4.5 \microns and 8 \microns images by overlaying a schematic of the geometry of the wide angle outflow with its inner boundary matching the outer extent of the jet and bow shocks.  The 8 \microns emission delineates the narrow jet traced by the entrained \h2 emission, while at  4.5 \micron, in addition to the jet, the wide angle outflow cavity is clearly visible.  The scattered light seen in the 3.6 \microns (Fig. 1)  and 4.5 \microns (Figs. 1 \& 4) delineates a sharp conical  boundary for the outflow cavity.  The projected opening angle of the cavity is $\sim$ 85\degs in the 3.6 \microns image which   is significantly large and
places it among those with very wide outflows (Velusamy \& Langer 1998; Arce \& Sargent
2006) and those with simultaneous presence of high velocity jets (Velusamy et al. 2007; Torrelles et al. 2010 and references therein). The wide angle outflow seen here in the scattered light closely resembles  the $^{13}$CO(1-0) low velocity lobes (Moro-Mart\'{i}n et al. 2001) with the blue shifted SW lobe approaching the observer.  However, the spatial resolution in the CO maps (8\arcsecs $\times$ 7\arcsec)  cannot resolve the sharp edges of the wide angle outflow cavity walls as seen more clearly  in the scattered light images with sub-arcsec resolution presented here.

 Poorly collimated flows can be due to extreme precession of the jet (e.g. Shepherd et al. 2000) and are indistinguishable from the apparent wide angle outflows.  In the Spitzer image, shown in Fig. 4(a), there is a ``second jet/outflow'' nearly perpendicular to the main north-south outflow as indicated by the east-west ellipse.  The presence of this secondary jet was first observed,  earlier in the H$_2$ 2.12 \microns image (Eisl\"{o}ffel et al. 1996).  The two brightest knots in the 4.5 \microns image were clearly detected in their H$_2$ 2.12\microns image while the eastern most knot is outside the 2.12 \microns image. In a later H$_2$ 2.12 \microns image (Ladd \& Hodapp, 1997) all three 4.5 \microns knots are detected.  The presence of two outflows nearly perpendicular to each other suggests the driving source in Cep E is a binary with a separation 4 –- 20 AU (0.005\arcsecs –- 0.03\arcsec),  and disk radius 1–10 AU (Terquem et al. 1999). A binary systems enhances the effects of precession of both the primary and secondary jets.  Eisl\"{o}ffel et al. interpret the wiggles in the loci of the H$_2$ 2.12 \microns knots as due to precession with a period of 400 yr and a pitch angle 4\deg. Binaries with  multiple  outflows may be  common; for example: HH111/HH121 (cf. Gredel \& Reipurth, 1993) and L1551 (cf. Rodr\'{i}guez et al. 1998).  Therefore we cannot rule out that the observed wide angle cavity in Cep E is the result of precessing jets. Indeed the combined effect of the precessing jet along with the geometry of the "infall envelope" (discussed in section 3.5) is an alternate scenario as the cause for the observed wide angle cavity.   In the absence of direct observational evidence on the outflow motions using high spatial resolution data of CO or other molecular tracers we cannot identify with certainty   the wide angle cavity detected in scattered light as wide angle outflow.

 In the jet driven outflow models the outflow emission near the protostar may be interpreted as the wings of the bowshocks extending towards the protostar (e.g. Raga et al. 2004).  However, in these models the opening angle of the  cavities near the protostar tends to be extremely narrow, in contrast to the wide angle observed in Cep E. Furthermore, the   emission in the the bow shock is distinctly different, predominantly \h2 emission, as seen in the IRAC images along the jet axis; but in the wide angle outflow region the \h2 emission is totally  absent as seen by the high contrast between the emissions in the 4.5 \microns and 8 \microns bands in Fig. 4. The lack of \h2 emission and the prevalence of scattered photospheric  light rule out the possibility that the wide angle outflow is  part of the jet driven bow shock.  Thus we can  constrain the extent of any jet precession to be strictly confined to the emission region in the 8 \microns image as shown by the schematic in Fig. 4.

  Whatever is the cause for the   wide angle cavity in Cep E (by precessing jets or by wide angle low velocity outflow), in Spitzer images it is observable only in the scattered light. We can clearly distinguish between   the jet and the wide angle cavity by using the  IRAC colors for scattered light in the wide angle cavity and \h2 emission in the narrow high velocity jets.  In Fig. 4a the   8 \microns emission (dominated by \h2 emission in the jets  and bow shocks) correlates well with the inner narrow jet features  in the 4.5 \microns image, but not with the biconical emission away from the jet.  Thus it is clear that biconical low surface emission is distinctly different from the jet emission and it is coming from the scattered light   in a wide angle cavity. In Fig. 4a the geometry of the  wide angle outflow   and the narrow high velocity  jets   in the northern lobe are shown schematically by conical boundaries. The emission in the region outside the inner cone  clearly represents the scattered light in the wide angle outflow and that from inside the inner cone boundary represents the jets and their associated bow shocks.   To study the spectral characteristics of the scattered photospheric  emission from the protostar in the wide angle outflow cavity we were able to separate the scattered emission from the jet emission by computing the flux densities   at each IRAC band integrated over the area between the inner and outer boundaries of the outflow cone as shown in Fig. 4a.    The scattered light flux density spectral plot is shown in  Fig. 4b.   This spectrum is distinctly different from those in Fig. 6a for the ``knots'' in the jets and bowshocks.  On the other hand,  between 4.5 \microns and 8 \microns it is suggestive of the  photospheric (at T=4300K) emission spectrum (shown in Fig. 9) obtained from the SED analysis (see section 3.4),  if we take into  consideration the effects of higher extinction at 3.6 \micron.   The  K-band extinctions are  in the range of 2.1 to 2.4 mag across the northern lobe and 1.4 mag for the southern lobe (Smith et al. 2003). The lower flux density at 3.6 \micron,  even after the  extinction correction  (using the near-IR to mid-IR extinction relations given by Indenbetouw et al.  2005;  Chapman et al. 2009),  would still be consistent with the examples of the SEDs for the scattered light which show   a dip near 3 \microns  due to the scattering albedo (Whitney et al. 2003b). A   comparison between the stellar and the scattered light spectra  will require detailed modeling including the grain properties and the outflow geometry (Whitney et al. 2003a \& b) as  presented for the bright scattered light emission from NIR to MIR  in the bipolar outflow cavities of the Class 0 object L1527 (Tobin et al. 2008). However, it is possible that, in addition to the complex inner structure generated by the outflow cavities, some  of the observed scattered light could result from the outer layers of the extended disk which forms from  infalling matter (Tobin, Hartman, \& Loinard, 2010).

\subsubsection{H$_2$ Jet Component}

Most of the
thermal energy associated with the shocks in the outflow is  radiated
away through the \h2 emission in the  rotational transitions; in the L1157 outflow, Nisini et al. (2010) estimate   about 50\% of the
total shock radiated energy  is in  \h2 emissions.
The similarity between   the jet/bowshock emissions along the jet axis in all the IRAC images is due to the fact that the mid IR jet emission is dominated by the \h2 pure  rotational transitions and each IRAC band contains several of the \h2 lines. (See Neufeld \& Yuan (2008) and Ybarra \& Lada (2009) for the details on their contributions to the band intensities.)  Below we list   the \h2 lines in each IRAC band and their upper state energy in degrees K, as a measure of the excitation required for their emission:\\
{\bf  3.6 \microns:} \h2 v = 1 -- 0 O(5)[6952K]; O(6)[7584K]; O(7)[8366K]; \h2 v = 0 -- 0 S(13)[17445K]\\
{\bf  4.5 \microns:} \h2 v = 0 -- 0 S(12)[15542K]; S(11)[13704K]; S(10)[11941K]; S(9) [10362K]\\
{\bf  5.8 \microns:} \h2 v = 0 -- 0 S(8) [8768K]; S(7)[7197K]; S(6) [5830K]\\
{\bf  8.0 \microns:} \h2 v = 0 -- 0 S(5) [4587K]; S(4)[3475K]\\
To demonstrate how these \h2 lines could be used to study the outflows,   Smith \& Rosen (2005)  show examples of synthetic images of \h2 emission in each IRAC band using model images obtained in hydrodynamical simulations of dense supersonic  molecular jets. In addition to the lines listed above, Smith \& Rosen (2005)  include in each IRAC band a few other  \h2 v = 1 -- 1 and \h2 v = 2 -- 1 transitions. The  3.6 \microns band is dominated by emission from vibrationally excited \h2 lines with higher mean excitation energies, while the other bands are dominated by rotational lines.

The \h2 emission in a few selected outflows have been well studied using the \h2 pure rotational
lines from S(0) to S(7) with IRS mapping (e.g. Neufeld et al. 2006; Nisini et al. 2010).  No IRS mapping data is available for Cep E, however,  Noriega-Crespo et al. (2004b)  used  the \h2 rotational transition lines in the  IRS spectrum at a single pointing near the brightest emission in the northern lobe  (see Fig. 6)  to estimate the mean excitation temperature (686 K)  and   a number density (6$\times$10$^4$ cm$^{-3}$). Moro-Mart\'{i}n et al. (2001) used the pure rotational H$_2$ 0-0 lines S(2) to S(7) intensities in 7 discrete positions  observed by  ISOCAM CVF   and their excitation diagrams, to model the temperatures and densities in the bowshocks.
The mid-IR \h2 emission  is  interpreted in terms of shocks produced by the high velocity atomic/ionic jets. High-density C-shock models can best  account for the brightness of the     \h2 emission (Smith et al. 2003; Dionatos et al. 2010).   In Cep E by comparison with C-type
shock models,   the \h2 line ratios observed in the IRS spectrum   give a mean shock velocity of 20 km s$^{-1}$ (Noriega-Crespo et al. 2004b).   Since the intensities of the \h2 lines within   each IRAC band are sensitive to a wide range of excitation conditions, the relative intensities in each band can be used as a diagnostic  of the physical conditions in the shocks.  As demonstrated by Ybarra \& Lada (2009) we can therefore, use the    IRAC colors:   [3.6] -- [4.5] {\it versus} [4.5] -- [5.8],  to study the  shocked H$_2$ at sufficiently high temperature (2000K -- 4000K) and neutral atomic hydrogen density (10$^3$ -- 10$^4$ cm$^{-3}$) in jets/outflows.

The 8 \microns image clearly delineates the outer boundary of the jets and their associated bowshock emissions   as shown by the schematic in Fig. 4. In all IRAC bands within this region   a number of jet features with resolved bowshock structures are identified along both north and south lobes.  Near-IR high resolution \h2 1–-0 S(1) imaging has shown that
the lobes consist of numerous emission knots (Eisl\"{o}ffel et al. 1996; Ladd \& Hodapp, 1997).    Smith et al. (2003) estimate  about 20 bow shocks, close to paraboloidal in
shape, are required by the model to explain the integrated
emission for each lobe. As indicated in the left panels in Figs. 5a \& 5b, in the IRAC 8 \microns image we identify at least 22 such resolved jet/bowshock components.   As seen in Fig. 1 the emissions in both lobes along the jet axis  in all four IRAC bands appear to be morphologically very similar.  Although all the knots identified at 8 \microns  (Fig. 5) have counterparts in all other bands, their relative intensities vary widely. Furthermore, in the 4.5 \microns image we could easily identify at least an additional seven knots (indicated as A1 to A7 in Fig.5) which appear  less prominently at 8 \micron.   To characterize fully the shock conditions in these \h2  knots and bowshock features and to bring  out their differences,   we constructed  the IRAC channel difference images (Figure 5): 4.5 \microns -- 3.6 \micron, 5.8\ microns -- 4.5 \micron, and 8  \microns -- 4.5 \micron. The difference image 4.5 \microns -- 3.6 \microns is more like  a continuum subtracted \h2 line emission (assuming the 3.6 \microns band behaves like a continuum filter with little \h2 line emission) and resembles the 2.12 \microns image (Eisl\"{o}ffel et al. 1996; Ladd \& Hodapp, 1997), tracing the brightest emission closer to the leading bowshock edges   on small scales, while the difference images,  5.8 \microns - 4.5 \microns  and 8  \microns -- 4.5 \micron, show   strong color differences determined by the  shock excitation along the wings, spatially away from their leading edges.

To obtain a more quantitative description of the knots identified in Fig. 5  we measured their flux densities  in each IRAC band by integrating over a 1.5\arcsecs $\times$ 1.5\arcsecs area and these are also shown as spectral plots in  Fig. 6a.  Though the images have sub-arcsec resolution, to obtain a more robust estimation of the flux densities we used a larger area of integration.      These spectra indicate that the excitation conditions vary widely among the knots: fourteen (including all seven selected by the 4.5 \microns peak emission)  have apparent spectral peaks at 4.5 \micron; ten have spectral  peaks between 5.8 and 8 \micron, while five have no apparent peak with fluxes increasing beyond 8 \micron. These differences are also clearly brought out in the difference images shown in Fig. 5. The extinction corrections will increase the fluxes at 3.6 and 4.5 microns relative to the others; however, the spectral shapes do not change significantly.   Our results of such strong 4.5 \microns  emission is consistent
with the    hydrodynamic simulations of Smith \&
Rosen (2005).  The knots with the highest excitation as inferred from the narrow spectral peaks at 4.5 \microns   (nos. K7, K6 and K2) are all located near the brightest tips of the  southern and northern lobes respectively. While the knots    K7 and K6  are among the brightest,  the knot    K2 is only marginally bright.  Except for these three, there is no other striking correlation between the  excitation conditions (spectral peak/shape) in the knots and their radial distance from the protostar or any particular age sequence.

The IRAC \h2 emission spectral data presented here will be useful to constrain the C-shock models for each jet feature along with those available in the near IR. We use the colors  [3.6] -- [4.5] and [4.5] -- [5.8] derived from the spectral data for each knot as   diagnostics to constrain the temperature and densities as discussed by Ybarra \& Lada (2009).  The  IRAC color-color plot for the H$_2$ knots in Cep E is shown   in   Fig. 6b.  The labels in Fig. 6b represent the color-color regime for each H$_2$ knot,  identified in Fig. 5.  The IRAC color-color plot for Cep E knots is remarkably similar to those obtained for HH54 (Ybarra \& Lada, 2009).  The majority of the knots, 17 out of 29, appear to have excitation conditions corresponding to temperatures 2500K to 4000K and densities 10$^3$ to 10$^4$ cm$^{-3}$.

The observed bowshock structures (leading edges and wings) projected on the sky  and the  subsequent interpretation of their shock layers emitting in different IRAC bands is critically dependent on the observer's viewing geometry with respect to the shock. Taking advantage of the sub-arcsec resolution in the deconvolved IRAC images we can  study in some greater detail the spatial characteristics in each image to resolve shock geometry. In Fig. 7  we show a blowup of the IRAC images near the tip of the NE and SW lobes.  We use the geometrical shape of the 4.5 \microns and 8  \microns emissions of the shock features combined with their relative displacements to ``infer''  the direction the shocks which is likely to be consistent with the schematic bowshock structure, for example as shown in Fig. 13 in Moro-Mart\'{i}n et al. (2001). In such cases we expect the atomic/ionic emission lines to originate at the tip of the bowshock (in the "Mach disk" and stagnation tip) where the J- shocks are dominant. But, the \h2 line emissions originate farther down along the bowshock wings where the shocks are oblique and more likely to be C-type. For a better understanding of the relationship between the structural complexity in the IRAC emissions and their colors with the atomic/ionic jet head, we plot in Figs. 7a \& 7b the atomic jet traced by the MIPS 24 \microns emission from the [Fe II] and [S I] lines (discussed below in section 3.3). The 24 \microns emission contour defines the location of the jet head (Mach disk/stagnation tip) which is driving the bowshock.  The high excitation \h2 lines in the 4.5 \microns band are likely to be relatively closer to the "tip" than the low excitation traced by 8 \microns band.  Schematically, in Figs. 7a \& 7b  the 4.5 \microns intensity contours overlaid on the 8 \microns grey scale delineate the bowshock closer to its leading edge (the tip) while the 8 \microns emission traces the projected spatial structure the bowshock wings away from the tip. The bright and dark bands seen across several knots in the 5.8\microns -- 4.5\microns  and 8 \microns -- 4.5\microns difference images (eg.  nos. 9, 10,12,  K4, K7, 21 in the right two panels in  Figs. 5a \& 5b) trace the projected separation between the bowshock tips and wings.  The geometry in two systems of knots consisting of nos. 3/5/6 (Fig. 5a and 7a) and K5/18/19 (Fig. 5b) respectively,    seems consistent with each one describing  a bowshock tip  with  wings on either side,   moving nearly in the plane of the sky. The two outermost arc-like emission features in the 4.5\microns emission (Fig. 7a) are also detected at   2.12 \microns (Ladd \& Hodapp, 1997). However, the outermost arc-like  emission at 8 \microns (Fig. 7a) is located between the two 4.5 \microns arcs,    behind the leading outer 4.5\microns arc and slightly ahead of the second 4.5\microns arc,  with  $\sim$ 1.5\arcsecs (1000 AU) separation between them. The 8 \microns arc does not seem consistent with bowshock wings but rather with a shocked layer between two successive leading shocks described by the 4.5\microns and 2.12\microns  arcs and originating in two successive ejections. It is interesting to note that the 4.5 \microns outermost arc has a flux density spectrum (K1 in Fig. 6a)  resembling that of the scattered light in the wide angle outflow cavity (Fig. 4b). It is, therefore, possible that this outermost arc indeed represents scattered light in the  cavity produced by the NE jet.  However, the continuum subtracted   2.12 \microns \h2 line emission  (Ladd \& Hodapp, 1997) strongly favors its identifications as a bowshock \h2 emission feature.

 The IRAC data for emissions near the SW tip as shown in Figs. 5b and 7b suggest a very complex system of oblique (C-type) shocks with possible twists in  the jet kinematics and geometry. The cartoon in Fig. 7b shows a plausible scenario for the C-type shocks, seen in projection, as traced by IRAC bands.    The  location of the jet head (Mach disk/stagnation tip) which is driving the bowshock is indicated by the 24 \microns emission contours.   The multiple C-type shocks suggested by the 4.5 \microns and 8 \microns emissions, surrounding the 24 \microns jet feature,  seem to be tracing different pieces of the bowshock and its wings,  expanding into   or out of the plane (but seen in projection on the sky), all  driven by   a single jet-head.  In Fig. 7b the long arrow marks the general direction of the atomic jet while the short arrows indicate oblique shocks moving in and out of the plane.  We can  interpret the relative outward and inward displacements of the 4.5  \microns emissions  with respect to the 8\microns for knots K7 and 21, respectively, as due to the viewing geometry of the observer and the C-type shock orientations.   As another example of the geometry of the bowshock with respect to the Mach disk,  in Fig. 7a we show the association between the 24 \microns atomic/ionic jet feature (\#2 in Fig. 8a)) and the bowshock traced by knots 4, 5 and 6 (identified as arc no.3 in Fig. 7a).

\subsection{Atomic Jets: MIPS 24 \microns Compact Sources}

In Fig. 8  we show an overlay of the  MIPS 24 \microns HiRes image (as contours) on the 4.5 \microns image.  At 24 \micron, in addition to the protostar, we detect at least three compact features coincident with the brightest emission in the bow shocks as traced by the \h2 emission in the 4.5 \microns and other IRAC bands.    These compact emissions at MIPS 24 \microns  coincide with the jet heads.   Although some enhanced   emission from the  collisionally
  heated dust in the shocks is possible, it is more likely that  the  24 \microns emission is
 tracing the hottest atomic/ionic gas in the bow shocks, possibly similar to that seen in the HH46/47 system (Velusamy et al. 2007).  In contrast to the morphology of the \h2 emission, which are bow shocks with extended long arcs,   the 24 \microns  emission is more compact and is confined to the tip of the bow shock coinciding with  the jet head. We can expect to see such differences if the   24 \microns emission is produced by   higher excitation (velocity) shocks
  arising closer to the apex of the jet as in the case of J-shocks, while the \h2 emission is produced further back in the low velocity C-shocks.
We can, therefore, interpret the 24 \microns emission to  represent that produced in the   J-shocks (ionic/atomic shocks)   from the bright [Fe II] line emissions at 24.51 and 25.98 \microns within the MIPS pass band, as observed in the case of HH46/47.   For one of the sources, \#1 in Fig. 8a, we can  use the IRS spectral data to substantiate the fact that the 24 \microns emission is indeed atomic.  In the previously published IRS spectrum (Fig. 6 in Noriega-Crespo et al. 2004b) of this feature we identify only the [S I] $^3P_1-^3P_2$ line at   24.249 \micron, but no [Fe II] line emissions.  In order to characterize the atomic emissions within the MIPS 24 \microns pass band, we replotted the spectrum using the  IRS data obtained from the {\it Spitzer} Archives.  This spectrum,  shown in Fig. 8b, is consistent with the published spectrum except for the baseline subtraction. The [S I] line first detected by Noriega-Crespo et al. (2004b) in this position in Cep E is clearly seen in this spectrum.  The [Fe II] line at  25.98 \microns and the \h2 S(0) at 28.22 \microns are also detected with relatively low intensities.  We use the intensities in this spectrum ([S I] at 0.82Jy and [Fe II] at 0.06 Jy) to estimate qualitatively the contribution to the 24 \microns pass band from the atomic spectral lines within the MIPS pass band as indicated in Fig. 6b.  The total integrated line emission corresponds to  0.64 Jy in a single IRS channel (0.17 \microns wide which is about 3\% of the total pass band) appearing at $\sim$ 60\% level of the pass band response.  We would then expect  this line emission to appear  as a 13 mJy continuum source in the MIPS 24 \microns image.    Thus, in the HiRes image (with 2\arcsecs spatial resolution) we expect it to have a brightness of $\sim$ 136 MJy sr$^{-1}$. This value is in reasonable agreement with the observed brightness of 103 MJy sr$^{-1}$, considering the uncertainties in estimating the equivalent intensity in the MIPS continuum band for a given channel intensity in the IRS spectrum.  We  therefore conclude that the 24 \microns emission in this jet feature in Cep E is truly atomic in origin, primarily from [S I].  Thus after   HiRes reprocessing, the {\it Spitzer} MIPS 24 \microns image is found to be an extremely useful diagnostic of the atomic component of the protostellar jets.

For the other two 24 \microns sources (\#2 and \#3 as marked by the short arrows in Fig. 8a)  seen along the jet emission, there are no matching IRS spectra to interpret the origin of their emission as atomic/ionic gas in the jet heads.   However in the case of \#3 strong [S II]$\lambda\lambda$6717/31 emission is observed by Ayala et al. (2000) in this position which coincides with  HH 377.  - Furthermore, [S II] and [Fe II] have very similar excitation energies  (Nisini et al. 2002) and therefore, we can expect a significant contribution from [S I] and/or [Fe II] to the MIPS 24 \microns emission.    Their location along the jet axis coincides well with the bright  emission knots/jet heads (inferred as arising from H$_2$) seen in the IRAC bands.  Also their positions coincide with the tip of the bow shocks. We expect to see [S I] and [Fe II] fine structure line
emission in the spectra of HH objects, which  show
the classical signatures of collisional excitation similar to those observed in the IRS spectra of supernova remnants like
IC 443 (Noriega-Crespo et al. 2009).   Therefore, by analogy to that identified in the jets of HH46/47 system (Velusamy et al. 2007), we can regard these 24 \microns emission features to have the same characteristics as the source \#1 discussed above.  Among the three, the source \#3 in the south lobe is the brightest (peak at 345 MJy sr$^{-1}$) which is also the brightest emission in the IRAC bands.  The second brightest source \#2 (with peak 145 MJy sr$^{-1}$) is located in the NE lobe, farther away than source \#1, but the IRAC emission here is weaker than for source \#1.   In the close vicinity of the sources \#1 and \#2 there are multiple bow shocks and jet heads in the IRAC bands. However the 24 \microns peaks are exclusively coincident with only one of the jet heads implying that not all bow shocks and jet heads have associated 24 \microns sources. In other words, not all jet heads have detectable atomic/ionic emission.   The two outermost  bow shocks in the northern lobe, which are clearly seen in the IRAC bands (see Fig. 7a), do not show any compact 24 \microns sources. Furthermore, these bow shocks do not show the bright jet heads as seen in the other cases.

As pointed out earlier, not all jet heads or knots with strong  \h2 emission in the IRAC bands have associated atomic/ionic counterparts.  The atomic/ionic emission appears to require higher excitation shocks than those producing the mid-IR \h2 lines.  Dionatos et al. (2010) discuss in detail  the physical conditions required for the mid-IR atomic/ionic and \h2 line emissions in these jet features.  They show that the low
velocity 10 -– 15 km s$^{-1}$ shock models that best fit the warm \h2
emission are unable to reproduce  the observed [Fe II] in HH211. It would be difficult for us to  quantify the differences in \h2 and atomic/ionic emission from the jet heads or knots in Cep E without appropriate shock models.  However, the association  between the 24 \microns atomic/ionic jet feature and the bow shock knots observed in IRAC bands can be   interpreted  in terms of the schematic bowshock  model as discussed above in section 3.2.2.   The 24 \microns feature  \#3 is associated with several knots K6, K7, 21 and 22   (in Fig. 5b and Fig. 7b) and   it can be identified as the Mach disk driving a system of oblique C-type shocks represented by these knots.   Similarly, the 24 \microns compact source  \#1 is associated with  the 4.5 \& 8 \microns knots no. 10, K3 and K4  respectively (Fig. 5a) and can be interpreted as the atomic/ionic jet driving the bow shock represented by these knots around it.  The 24 \microns atomic/ionic jet feature \#2 brings out even more clearly its association with the bow shock and its wings observed in IRAC bands.     This feature is associated with the knots 4, 5 \& 6, and the diffuse emission along the bowshock wings   (arc no. 3 as shown in Fig. 7a).

So far, we have IRS spectral data for only  two 24 \microns compact sources associated with the atomic jet head (knots), but they  seem to have very different atomic spectral characteristics.  The jet head in the HH46/47 system shows a strong [Fe II] line while that in Cep E  shows a strong [S I] line and very weak [Fe II]. Such differences  suggest that the physical conditions in the jet heads or knots can vary widely.   The IRS spectral line intensities in the jet heads or knots, similar to those in Cep E  and HH46/47, in  HH211 (Dionatos et al. 2010),   HH54, HH 7-11 (Neufeld et al. 2006), BHR71,
L1157, L1448, NGC 2071, and VLA 1623 (Neufeld et al. 2009),   show a wide variation in the intensity ratio  [Fe II]26 \microns to [S I]25 \micron.  Such variations can be interpreted as electron density and temperature effects on the ionization of Fe, as well as differences in the collisional  excitation. Neufeld et al. (2009) find  evidence for some spatial segregation between   the [S I] and [Fe II] emitting regions,    suggesting that [Fe II] is tracing the   faster, dissociative shocks.   Thus the differences between  the 24 \microns features in Cep E and HH46/47 are due to  their  shock conditions, namely shocks in Cep E being less dissociative than in HH46/47.

 The ISO-LWS spectra of both the north and south lobes in Cep E detected strong [O I] 63 \microns and [C II] 158\microns emissions (Moro-Mart\'{i}n et al. 2001), which are  the most efficient coolants in low velocity dissociative shocks.   The low intensity features in the deconvolved MIPS 70 \microns image (Fig. 2) are very intriguing, especially since they lie along the jet/outflow lobes in Cep E. Of special interest to us is the fact that the  MIPS 70 \microns band includes the [O I] 63 \microns line near the end of its passband at $\sim$ 75\% response.  In  the ISO LWS spectra the [O I] 63 \microns line was detected as the   brightest line emission with intensities  of 9.3 and 12 $\times$ 10$^{-19}$ W cm$^{-2}$ at the NE and SW lobes respectively. Such emission would be observed in the MIPS 70 \microns image as continuum sources with total flux densities of 0.58 Jy and 0.75 Jy respectively for the NE and SW lobes which are higher by a factor 2 to 3 than the features in Fig. 2. These feature are at intensities at a level of 0.1\% of the peak (protostar) emission.   The SW feature is the brighter of the two and in the 70 \microns Point Response Function (PRF) it coincides with an enhanced sidelobe in the Airy ring,  at $\sim$ 6.3\% level.    We believe the MIPS 70 \microns data does not warrant such high dynamic range ($>$ 1000) in the HiRes processed image.  These features appear to be residuals of the Airy ring.  Furthermore, their locations do not seem to correlate with any of the features observed in 24 \microns or the IRAC  bands.

\subsection{Photometric SED Analysis of  Cep E Protostellar Emission}

We obtained the {\it Spitzer} IRAC and MIPS photometry  using the mosaic
images.  For the IRAC data we used the artifact corrected BCD's (cbcd
images) and Imasks (bimsk), with the standard overlap and mosaic
modules, to create the mosaic images for each of the IRAC
bands. The mosaics were done using a pixratio=2, which results in a
pixel scale of 0.6\arcsecs per mosaic pixel.  IRAC photometry was
obtained by using the {\sc phot} task in IRAF. An aperture radius of 4
pixels (2.4\arcsec), sky annulus between 4 and 8 pixels, was used to
obtain the aperture photometry. For the 3.6 \micron,
 4.5 \micron, 5.8 \microns and 8 \microns  bands we used zero point magnitudes of  18.59, 18.09, 17.48 and 16.7
respectively, and corresponding aperture
corrections of 1.213, 1.234, 1.379 and 1.584.

The {\it Spitzer} MIPS BCD data in the 24 \microns and 70 \microns bands were
 used to create the basic mosaics using the relevant templates in the MOPEX
 software. Subsequently, photometry was obtained by using the
 appropriate default MOPEX/APEX single-frame templates for each
 band. The recommendations of the {\it Spitzer} data analysis cookbook
 recipe 24 were
 followed.\footnote{(http://ssc.{\it Spitzer}.caltech.edu/dataanalysistools/cookbook/29/)}
We note that our MIPS photometry agrees with the photometry derived
by Noriega-Crespo et al. (2004b)  to better than 5\%.

We constructed the spectral energy distribution of the Cep E protostar
by combining our {\it Spitzer} photometry with other photometry from the
literature. Chini et al. (2001) have studied this
protostar at millimeter wavelengths, using the  fluxes at 450 and 850 \microns obtained with   SCUBA on the JCMT
and at 1300 \microns with the IRAM telescope.   The aperture sizes used to derive the
fluxes are all larger than 10\arcsec, which is larger than the
aperture used to obtain the MIPS 70 \microns photometry. In order to
 model the protostar using the SED fitting tool by
Robitaille et al. (2007), it is essential to classify the available data
as ``data points'' and ``upper limits''. The data points are used to
obtain the best model fitting and computing the $\chi^2$ factor, whereas
the upper limits serve in constraining the fits. We note that the
protostar appears as a point source in the {\it Spitzer}-IRAC and MIPS
bands (Fig. 3). Therefore, we use the IRAC and MIPS photometry as data points
in subsequent modeling. The millimeter fluxes from Chini et al. (2001)
are obtained from apertures larger than 10\arcsec\, and could be
contaminated by the outflow and surrounding cloud. Therefore, we use
the millimeter data points as upper limits. We also use the IRAS
12 \micron\, flux (obtained from an aperture of 30\arcsec\,) as an
upper limit. The protostar is not detected in the 2MASS K band,
therefore, an upper limit of 16\,mag in the K band was used to constrain
the SED.

The SED, characterized by these data points and upper limits, is
modeled using the SED fitting tool (Robitaille et al. 2007). We note
that the actual photometry has an accuracy closer to 1\%-2\%. However, we  assumed photometric uncertainties of 10\% as a
conservative limit to minimize the risk of including biases in the modeling.   The SED
fitting tool requires a range of values for the distance and interstellar extinction   to
the source, in order to scale the models in the grid to the observed
data. The  lower and upper limits to the distance were set at 0.65 and 0.8 kpc,
respectively. The interstellar extinction was set to vary between 2
and 8 visual magnitudes.

The observed SED is well represented by a YSO model in the Class 0/I
evolutionary stage. Fig. 7 shows the models fitted to the SED. The
detailed physical parameters of the young stellar object system
derived from the SED fitting are listed in Table  2. The errors quoted
in Table 2 correspond to the spread in the parameters of the models
shown using the grey lines in Fig. 7. Which physical parameters are
most reliable depends on what kind of model best describes the
observed SED. This philosophy is different from other interpretations
of YSO's based on colors or spectral indices alone (for details see:
Smith et al. 2010; Gramajo et al, 2010). In a Class 0/I YSO model,  it is the envelope
properties, listed in Table 3, that are better constrained, in contrast to the disk properties,
because the disk is deeply buried inside the envelope. The envelope
mass for the Cep E protostar is $\sim$ 7M$_\odot$, with an accretion rate of
$\sim$ 10$^{-4}$\Msun yr$^{-1}$. The accreting photosphere embedded
inside this envelope is currently estimated to have a mass of
$\sim$ 3\Msun and a disk mass  $\sim$ 10$^{-2}$M$_\odot$. Together, these
values support the conjecture that Cep E will become an intermediate
mass star of $\sim$ 4\Msun (Froebrich et al. 2003; Froebrich, 2005) at the end of star
formation. The wide angle flow detected in this work may well be the
signature of rapid accretion and radiation driven wind from the
envelope, which is more appropriate for intermediate mass stars than
low mass stars. The inclination angle of the source, obtained by model
fitting, is $\sim$ 42\deg$\pm$17\deg. Our model fitted value agrees, within the uncertainties, with the inclination of 57\deg$\pm$7\deg\ derived from the radial velocity and proper motion (Smith et al. 2003).
\subsection{Infall Envelope as an IR Absorption Band}

Our SED analysis  show the existence of an arcmin size dense envelope with 7\Msun and size 20,000AU (see Table 3).  Such large scale protostellar infall envelopes perpendicular to the outflow are common in Class 0 protostars (cf. Velusamy \& Langer 1998; Chang et al. 2010).  In the Class 0 protostar L1157, also located in the Cepheus flare, an extended $\sim$ 2\arcmins flattened envelope perpendicular to the outflow was seen in the   IRAC  band at 8 \microns  in absorption against the background interstellar PAH emission (Looney et al. 2007; Tobin et al. 2010). Stutz et al. (2008) detected  a similar    extinction feature (which they refer to   as a shadow) around  the protostar in the IRAC 8 \microns image of B335 consistent with the flattened molecular core.    In Fig. 8  we show the mosaic images at 3.6 \microns  and  8 \microns of a larger region than shown in Fig. 1.  To the south of Cep E we can see  a large interstellar emission cloud possibly rich in PAH emission,  since it is not seen in the 3.6 \microns image shown for comparison in Fig. 8.    To the north of this cloud emission, on both sides, to the east and the west of the Cep E jet/outflow region, we see patches of 8 \microns absorption of this PAH background.    Two patches (indicated by the arrows in Fig. 8b) are seen on either side closest to the protostar. It seems likely that these are the ends of a continuous absorption band which is  masked in the middle by the protostar and its bright ``side lobes'' (Airy rings) around it.  It is possible that these absorption features are similar to those seen in L1157 by Looney et al. (2007) but is seen less clearly here because of the confusion by the emission in the sidelobes of the protostar.

Though the HiRes deconvolved image (Fig. 1) is free from the sidelobe contamination from bright sources in the image, we cannot use it for tracing the absorption feature. The HiRes processed images do not contain the background emission to trace this absorption features.  Optimal performance HiRes   requires that the background emission is fully subtracted out in each BCD prior to applying the deconvolution. Furthermore, the positivity criteria implicit in the deconvolution algorithm makes it insensitive to  negative intensities.  However we can still use the HiRes deconvolution on the BCDs with  the full background emission present and remove the sidelobes, but sacrificing the full resolution enhancement. For this purpose we stop the HiRes processing with fewer  than 10 iterations instead of 50 iterations used for optimal processing.  In Fig 8c we show the results of HiRes deconvolution    after 10 iterations  on the data with the background present. After the removal of the sidelobes  the protostellar envelope is traced more clearly by the 8 \microns absorption extending closer to the protostar.  We interpret this absorption feature as the  extended protostellar envelope.    We estimate an angular size of about 35\arcsec, corresponding to 22,000 AU which is in good agreement with that inferred from the SED analysis.   It is most likely that this absorption feature is the infall  envelope   which is feeding  the disk and the forming star in Cep E. Indeed, Gregersen et al. (2000) have detected  the infall signature in Cep E using the velocity asymmetries observed in HCO$^+$(3-2) with a 26\arcsecs beam.  A scenario with an extended infall region is particularly interesting in view of the simultaneous presence of collimated jets and wide angle outflow in Cep E. In the recent models by Machida et al. (2008) the high velocity jets and the low velocity wide angle outflow are driven by two different components, namely, the  accretion in the compact protostellar disk driving the jets and the accretion from the extended infall envelope driving the wide angle outflow.  Thus, along with the other examples of B335  and L1157, we show that the 8 \microns extinction   is a useful diagnostic of the protostellar envelopes.

\subsection{Simultaneous Jets and Wide Angle Outflows}

In some protostellar objects both wide angle outflows and collimated jets are seen, while in others only one is
observed. These observations raise a number of important questions about jets and outflows, which directly impact our understanding of their role in star and disk formation.  What is the origin of the wide-angle outflows and collimated jets and how do they evolve with time?  To date, jets and outflows from a protostar have various morphological and kinematical properties such that they cannot be explained by a single-class model.      The centrifugal force originates from the rotation of the circumstellar accretion disk and the jets are ultimately powered by the infall of material, with the rotation and magnetic fields playing a crucial role.  The magnetocentrifugal origin of jets and their launch from the magnetized accretion disk of the protostar (cf. Ouyed \& Pudritz 1997) are generally accepted, although the detailed mechanism is under debate.  The wide-angle outflow may also be jet driven (cf. Raga \& Cabrit 1993, Ostriker et al. 2001) or wind driven (cf. Shu et al. 2000).  In the jet-driven bow shock model, an episodic variation in jet velocity produces an internal bow shock driving an internal shell, in addition to the terminal shock.  Poorly collimated flows seen as wide angle outflows can be due to extreme precession of the jet (Shepherd et al. 2000) and indistinguishable from wind driven wide angle outflows.  Other explanations include a turbulent jet model (cf. Cant\'{o} \& Raga 1991; Watson et al. 2004), and a circulation model, involving radiation and magnetocentrifugal acceleration and collimation producing heated pressure-driven outflows (cf. Lery 2003, Combet et al. 2006).

In a unified model for bipolar jets and outflows Shang et al. (2006) incorporate the essential features of the primary X-wind, which is   driven magnetocentrifugally from close to the protostar   in the interface
between its magnetosphere   and the associated circumstellar disk. This primary
wind has an angle-dependent density distribution, with a dense axial jet surrounded by a
more tenuous wide-angle wind. The resulting
structure shows two prominent dense features: a shell of mostly
swept-up ambient material and a jet along the axis that is the
densest part of the primary wind. The shell can be either well-collimated,
as observed for the class of jet-like molecular outflows,
or wide open, as in the classical molecular outflows.  The morphology of  jets and  outflow is shaped
to a large extent by the ambient mass distribution in  the collapsing
envelope and thus jet-like and wide angle   outflows are  unified only
in an evolutionary sequence.
In the disk-wind model  simulations of collapsing, rotating, magnetized Bonnor-Ebert spheres of  molecular cores, the highly collimated wind is driven magnetocentrifugally from a
wide range of circumstellar disk radii, surrounded by a wide-angle wind driven by toroidal
magnetic pressure  (e.g.  Banerjee \& Pudritz 2006).
In another  unified model    Machida et al. (2008)   explain the simultaneous occurrences of both the jets and wide angle outflows using the results from their calculation of cloud evolution from a molecular cloud core to a stellar core, starting with a Bonner-Ebert isothermal core rotating in a uniform magnetic field.   They find two distinct flows, wide angle low velocity and narrow high-velocity flows driven by two different components: the collapsing core and the protostar, respectively.   These  two distinct flows have  different degrees of collimation and velocities: the low-velocity flow (i.e.  molecular outflow) has a wide
opening angle, while the high-velocity flow (i.e., optical jet) has
a well-collimated structure. This collimation is caused by both the configuration
of the magnetic field lines around the drivers and their
driving mechanisms.  The low velocity
wide-angle outflow, mainly driven by the magnetocentrifugal wind
mechanism (disk wind), and
guided by hourglass-like field lines; and the fast highly-collimated outflow driven by magnetic
pressure and guided by straight field lines near the protostar.

The results of the unified model by   Machida et al. (2008), as seen in  their schematic (Fig. 15 in their paper), are remarkably similar to the simultaneously present narrow jet and wide angle outflow   in Cep E we derive from {\it Spitzer} observations.   This model also appears to be   viable   for the jet/outflow seen in several other objects such as in B5-IRS1 in which the wide-angle CO outflow (Velusamy and Langer 1998), and the pc scale HH flows (Yu et al. 1999)    are likely be driven by the extended infall and the compact disk, respectively.   HH46/HH47 (Velusamy et al. 2007), CepA (Torrelles et al. 2011), and HH 211 (Hirano et al. 2006)  are a few other examples of simultaneous jets and wide angle outflows in   protostars.  In outflows of  L1448-mm and IRAS 04166+2706, Tafalla et al. (2010) observe evidence for the simultaneous presence of multiple components consisting of a wide angle slow wind (identified by the line-wings)  and collimated flows (identified by the extremely high velocity (EHV) component).    A high angular resolution CO(J = 2-1) map of the EHV component in I04166 is well explained as resulting from a series of internal working surfaces traveling along a collimated jet (Santiago-García et al. 2009).  The large scale morphologies  observed in these objects are broadly consistent with  any of the above models for the  multiple jet/outflow components.   Observations with much higher  spatial resolution  with sufficient sensitivities will be needed to probe the launching region of the primary winds on a scale of a few AU of the protostar to discriminate between various models of jets and outflows. Furthermore   a large  sample of fully mapped  jets and outflows will lead to a better understanding of the simultaneously present jets and wide angle outflow sources in the context of the evolutionary sequences of protostars.

\section{CONCLUSIONS}
By combining the high sensitivity in the  {\it Spitzer} images and the  reprocessing with HiRes  we find new jet and outflow features in Cep E in addition to those previously known; the new results are: (i) wide angle outflow seen in the scattered light; (ii) morphological details of at least 29  jet driven bow shocks and jet heads or knots; (iii) three compact features in 24 \microns continuum image identified as atomic/ionic line emission coincident with the  jet heads; (iv) a flattened arcmin size protostellar envelope seen  against the interstellar background PAH emission  as an absorption band seen  across the protostar at 8 \micron. We demonstrate that by separating the protostellar photospheric scattered emission in the wide angle cavity from the jet emission, we can study directly the scattered light spectrum. The simultaneous presence of collimated jets and wide angle outflow in Cep E, as shown here, is consistent with unified jet-outflow models such as by Machida et al. (2008) in which   the  accretion in the compact protostellar disk drives the high velocity jets and the accretion from the extended infall envelope drives the wide angle outflows. We have obtained the \h2 emission line spectra  as observed in all IRAC bands for all 29 knots and use their IRAC colors as probes of  the temperature and density  in the jets and bowshocks. These spectra are useful as diagnostics of the C-type shock  excitation of pure rotational transitions of \h2 and a few \h2 vibrational emissions. A shock excitation model template of IRAC  intensity spectra  taking into account all  \h2 transitions and the spectral response of  each IRAC band will be a useful resource for  interpreting  the vast {\it Spitzer} data base available to study  protostellar jets and  outflows. Detailed modeling of the individual shocks  will help retrace the history of episodic jet activity and the associated accretion on to the protostar.

Our results for Cep E, which utilize the HiRes algorithm,  show the potential of {\it Spitzer} data to study the entrained molecular \h2 in the high velocity jets as well as the wide angle outflow cavities  associated with the slower winds often traced by CO. Our approach for Cep E builds on the earlier work of  Velusamy et al. (2007), who identified simultaneous jets and wide angle outflows in the young protostar  HH46/HH47.  In addition to the {\it Spitzer} data,    the recent mm and sub-mm spectral line observations of outflows and jets from ground (e.g. Tafalla et al. 2010; Santiago-García et al. 2009) and from {\it Herschel} (e.g.  Codella et al. 2010; Lefloch et al. 2010),   will offer new clues to the nature of the EHV gas and its relation to the low velocity outflow shells.   Though a few selected outflows have been well studied in \h2 emission using all \h2 pure rotational
lines from S(0) to S(7) with IRS (e.g. Neufeld et al. 2006; Nisini et al. 2010) the IRAC images available in the {\it Spitzer} archives have the potential to study the \h2 emission in a large sample of outflow sources.  Quantitative analysis of IRAC colors using statistical equilibrium estimates such as those of Ybarra and Lada (2009), demonstrate the value of Spitzer images to probe the thermal structure of the shocked gas without the need of using spectroscopic data.  Thus, the {\it Spitzer} capability to study these features in \h2 and scattered light, as described here, will be complementary to the mm/sub-mm molecular data, thus leading to a better understanding of the origin of the atomic jets and the fast and slow  molecular outflows,  and their  subsequent  manifestation in the  physical and chemical characteristics as observed in the resulting shocks.    The {\it Spitzer} \h2 molecular outflow data is even more relevant now,  in the light of recent observations (cf. Tafalla et al. 2010) of the EHV component (in addition to the wing components) in a suite of complex molecules such as CO, SiO, SO, CS, HC$_3$N, HCO$^+$, H$_2$CO, HCN, H$_2$O and CH$_3$OH, that can be used to model  the shock chemistry and excitation conditions.
 \acknowledgments
We thank  Dirk Froebrich for useful discussions.  We thank the referee,  Alberto Noriega-Crespo, whose critical comments helped us to present a more complete and comprehensive picture of Cep E; in particular, for pointing out the presence of a second jet and suggesting using the IRAC colors.  The
research described in this paper was carried out at the Jet
Propulsion Laboratory, California Institute of Technology, under a
contract with the National Aeronautics and Space Administration.  M. S. N. Kumar is
supported by a Ci\^{e}ncia 2007 contract, funded by FCT/MCTES (Portugal) and
POPH/FSE (EC).


\clearpage

\begin{table}[htbp]
\caption{Observed Protostar parameters from the HiRes maps}
\begin{tabular}{cccc}
\hline\hline
Band & Peak &size  & Flux density  \\
\microns & MJy sr$^{-1}$ & \arcsecs $\times$ \arcsecs & Jy \\
\hline
3.6 & 355 & 0.83 $\times$ 0.72 & 0.0055 \\
4.5 & 1877 & 0.73 $\times$ 1.00 & 0.0358 \\
5.8 & 6465 & 0.71 $\times$ 0.91  & 0.111 \\
8.0 & 10964 & 0.78 $\times$ 0.85 & 0.196 \\
24  &47597 & 1.97 $\times$ 2.06 & 5.15\\
70 & 32714 & 5.23 $\times$ 6.03 & 26.4 \\
\hline
\end{tabular}
\\

\caption{Results from SED modeling: Protostar}
\begin{tabular}{cccc}
\hline\hline
Age & M$_*$ & R$_*$ & T$_*$ \\
log(yr) & log(M$_\odot$) & log(R$_\odot$) & log($T_{\odot}$) \\
\hline
4.33$\pm$0.1 & 0.51$\pm$0.01 & 1.24$\pm$0.01 & 3.64$\pm$0.00 \\
\hline
\end{tabular}
\\
\caption{Results from SED modeling: Disk and Envelope}

\begin{tabular}{ccccccc}
\hline
\hline
$\dot{M}_{env}$ & R$_{Max}^{env}$  & M$_{disk}$ & R$_{Max}^{d}$  & $\dot{M}_{disk}$ & incl &  M$_{env}$ \\
 log(M$_\odot$$yr^{-1}$) & log(AU)  & log(M$_\odot$) & log(AU)  & log(M$_\odot$$yr^{-1}$) & \degr & log(M$_\odot$) \\
\hline
-3.85$\pm$0.14 & 4.07$\pm$0.09   & -2.17$\pm$0.73 & 1.82$\pm$0.39  &  -7.53$\pm$0.52 & 42$\pm$17 & 0.87$\pm$0.28 \\
\hline
\end{tabular}

\label{}
\end{table}

 \clearpage
\begin{figure}[t]
\includegraphics  [ scale=0.65, angle=-90]{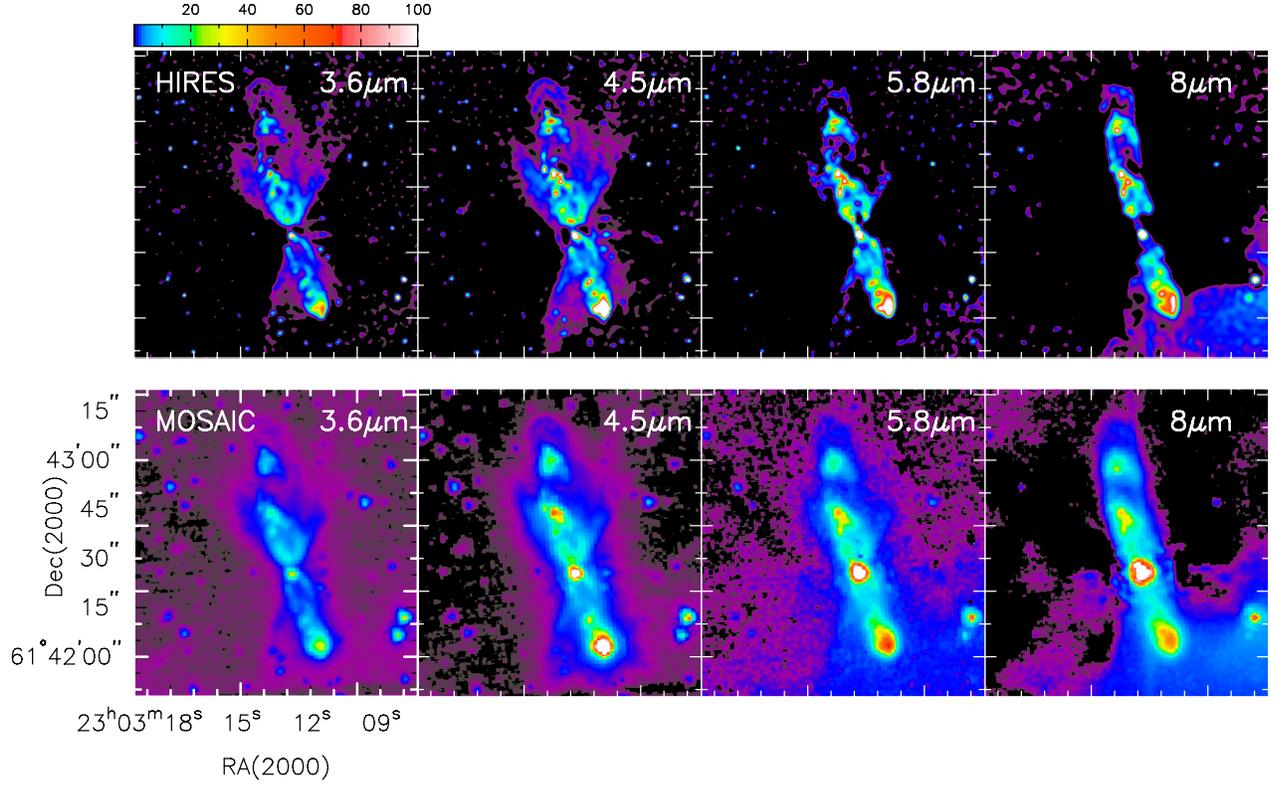}
\caption{Mosaic (lower) and HiRes deconvolved (upper)  images in the IRAC bands. In all images the  color stretch and the square root scaling used are designed to bring out the low surface emissions and thus the brightest emission is saturated.  It is readily evident that  the image enhancement in the HiRes images improves the  visualization of the morphologies of jets and outflows by virtue of the increase in the resolution to sub-arcsec scale, and by removal of the confusion from the ``side-lobes''. \label{irac}}
\end{figure}
\begin{figure}
\includegraphics  [ angle=-90]{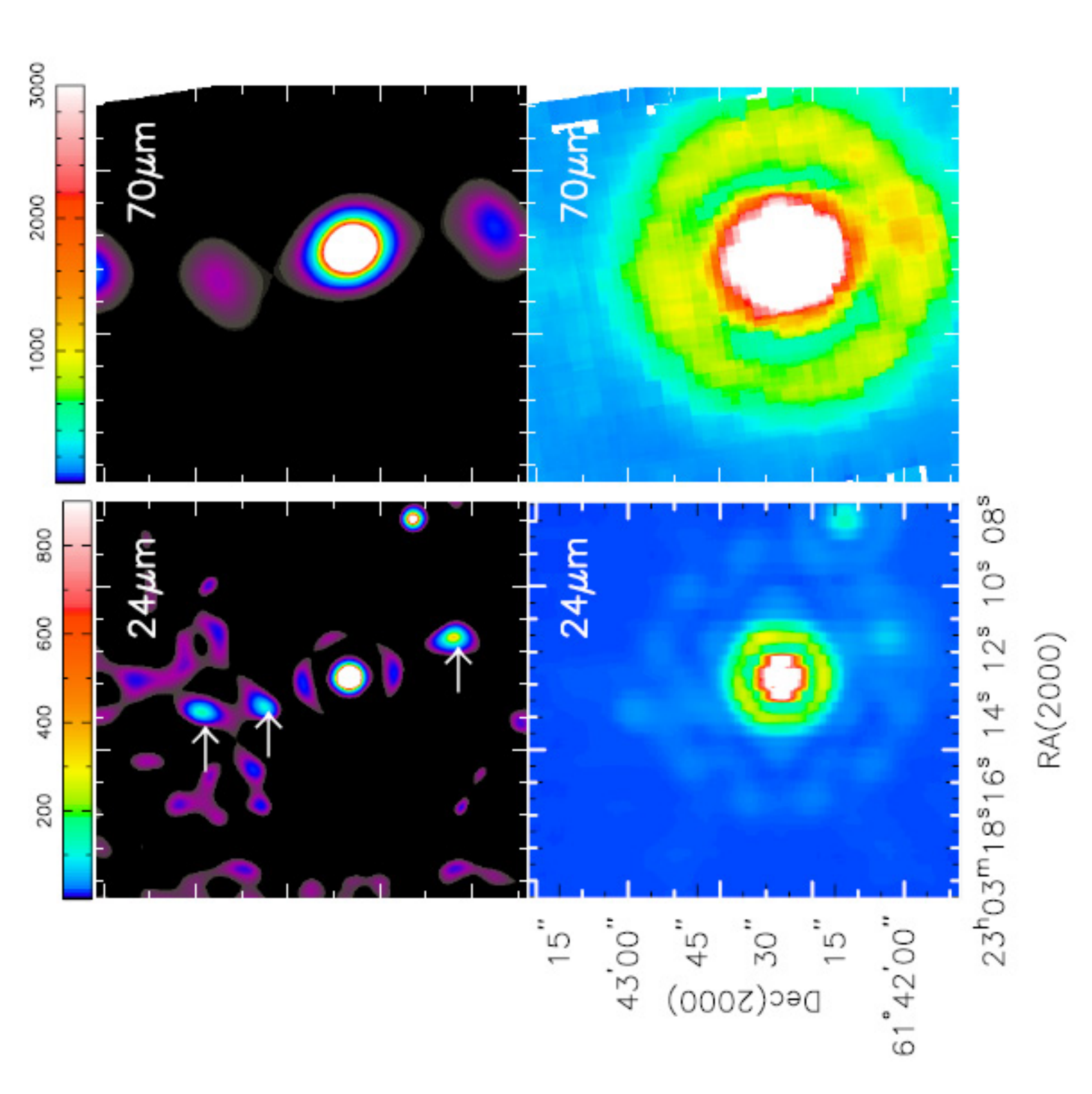} \caption{  Mosaic (lower) and HiRes deconvolved (upper)  images in the MIPS bands. The arrows in the 24 \microns HiRes image indicate the features which are distinctly visible  only after the HiRes processing and  were formerly completely masked by the ``side-lobes'' in the mosaic image.}
\end{figure}

\begin{figure}
\includegraphics  [angle=-90, scale=0.65] {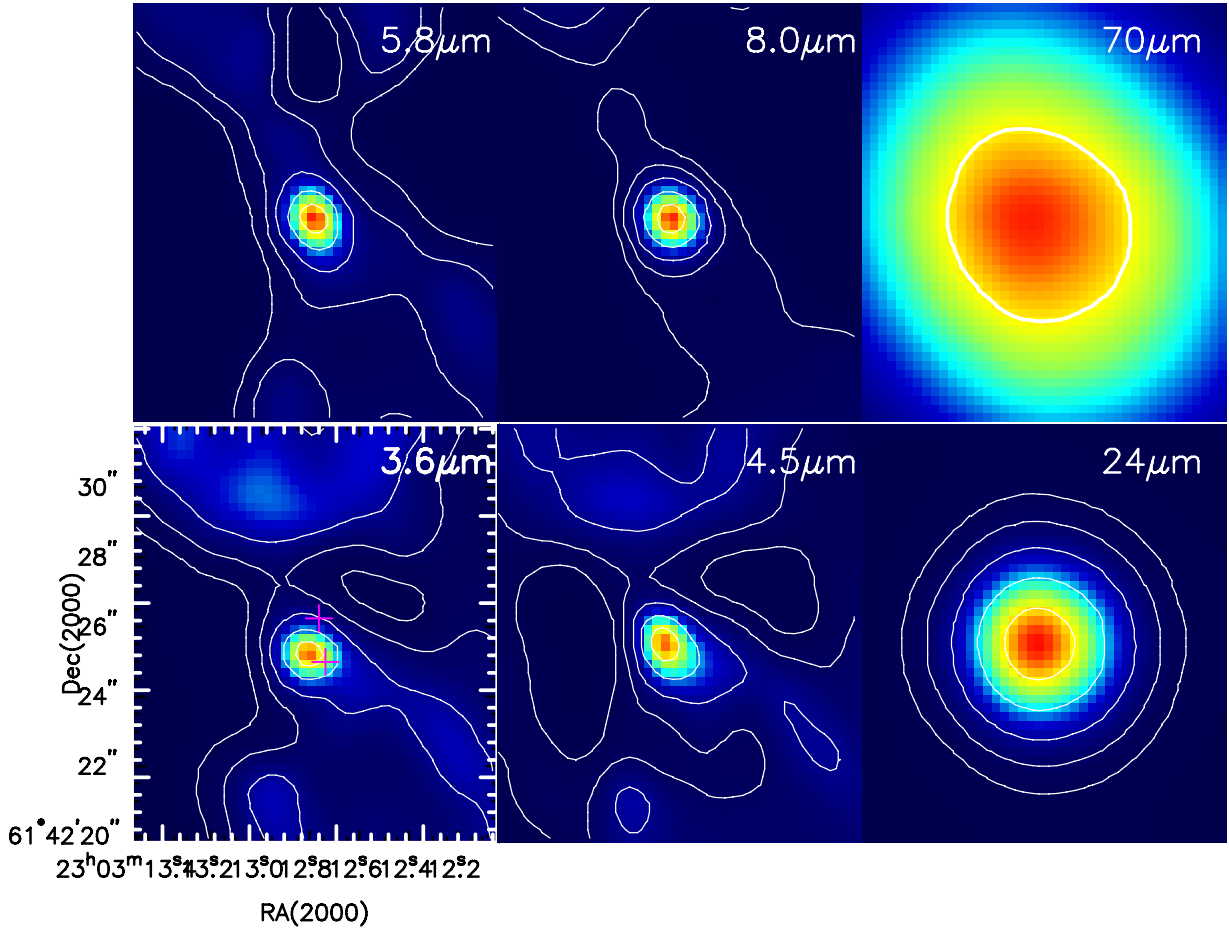}
\caption{ HiRes maps of the inner 10\arcsecs region around the Cep E protostar.  The contour levels are at 0.01\%, 0.1\%, 1\%, 10\% and 50\%  of the peak intensity (listed in Table 1) at each band. At 70 \microns only the contour at 50\% peak intensity is plotted.  Note the extremely low surface brightness of the outflow-jet features in all IRAC bands, delineated by the   low-level contours at intensities below a few percent of the peak at the protostar.  The crosses in the 3.6 \microns map represent the positions of the mm double source (from Moro-Mart\'{i}n et al. 2001).  \label{protostar}}
\end{figure}
\begin{figure}
\includegraphics  [scale=0.70] {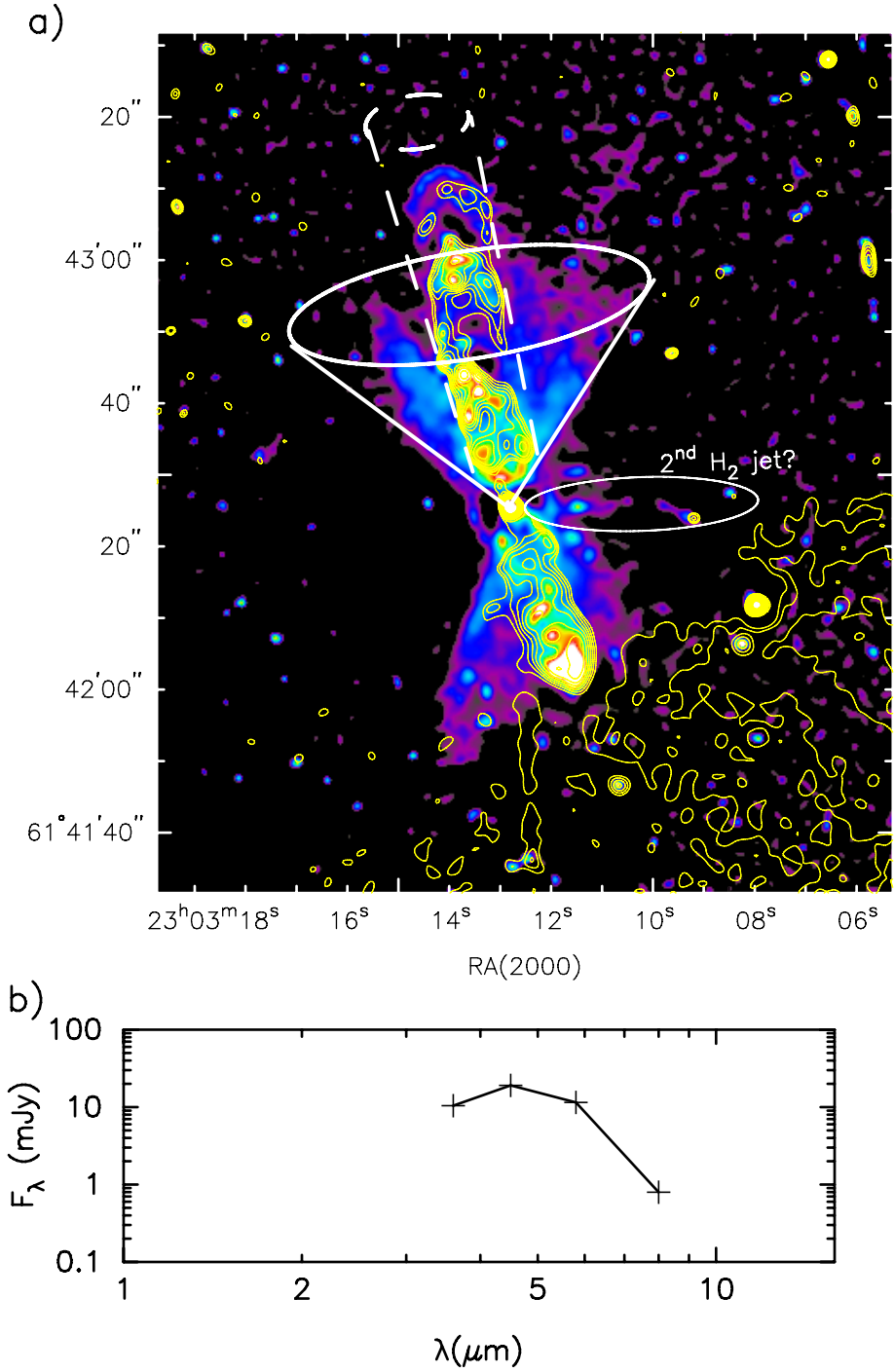}
\caption{ (a) The 8 \microns emission (intensity contours) tracing the jet overlaid on the 4.5 \microns image which traces both the jet and the wide angle outflow in scattered light.  The  contour  intensities are at  1, 2, 4, 8, 16, 32, 64, 128 and 256   MJy sr$^{-1}$ (peak intensity is 10,964 MJy sr$^{-1}$).   We show a schematic of the bi-conical wide angle outflow (solid lines) and the narrow jets and associated bowshocks (broken lines) in the northern lobe. The  plausibility of a second H$_2$ molecular jet/outflow (see text)  is indicated.     (b)  Spectral plot of relative intensities of the  scattered  light  integrated over the area between the boundaries of the  wide angle outflow and the narrow   jet as shown  in the upper panel.
\label{wideangle}}
\end{figure}
\begin{figure}
\includegraphics  [angle=-90, scale=0.55] {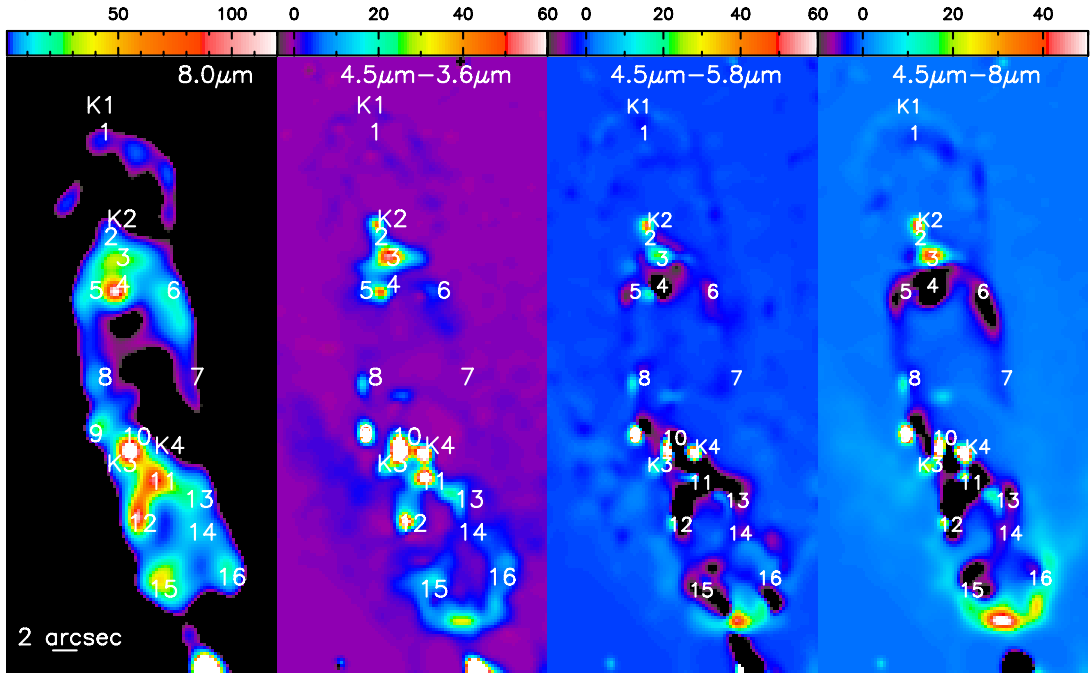}
\includegraphics  [angle=-90, scale=0.55] {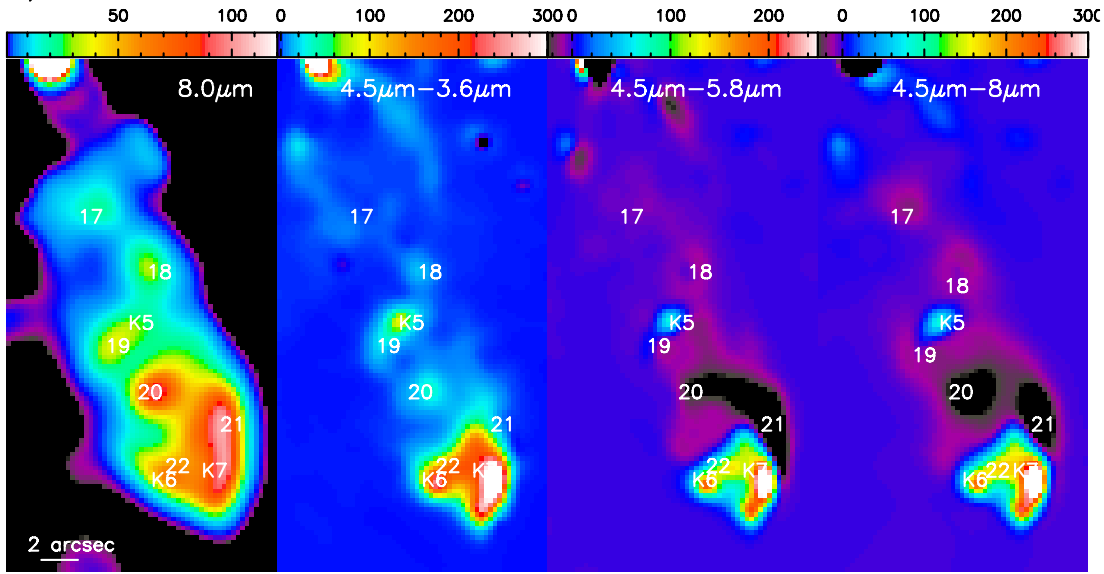}
\caption{The 8 \microns image (left panel), and difference images between IRAC bands (right three panels) showing the spatial variation of excitation across the  bowshocks inferred from the colors: (a) NE lobe; (b) SW lobe.   In \h2 a total of   29 jet features, ``knots'',  are identified as marked on the images in all panels.   The numbered labels 1 to 22 represent the knots identified by the peak emissions in the 8 \microns image.  The knots identified by K1 to K7 represent those less prominent at 8 \microns but are bright at 4.5 \micron.     \label{knots}}
\end{figure}
\begin{figure}
\includegraphics  [ angle=-90,scale=0.55] {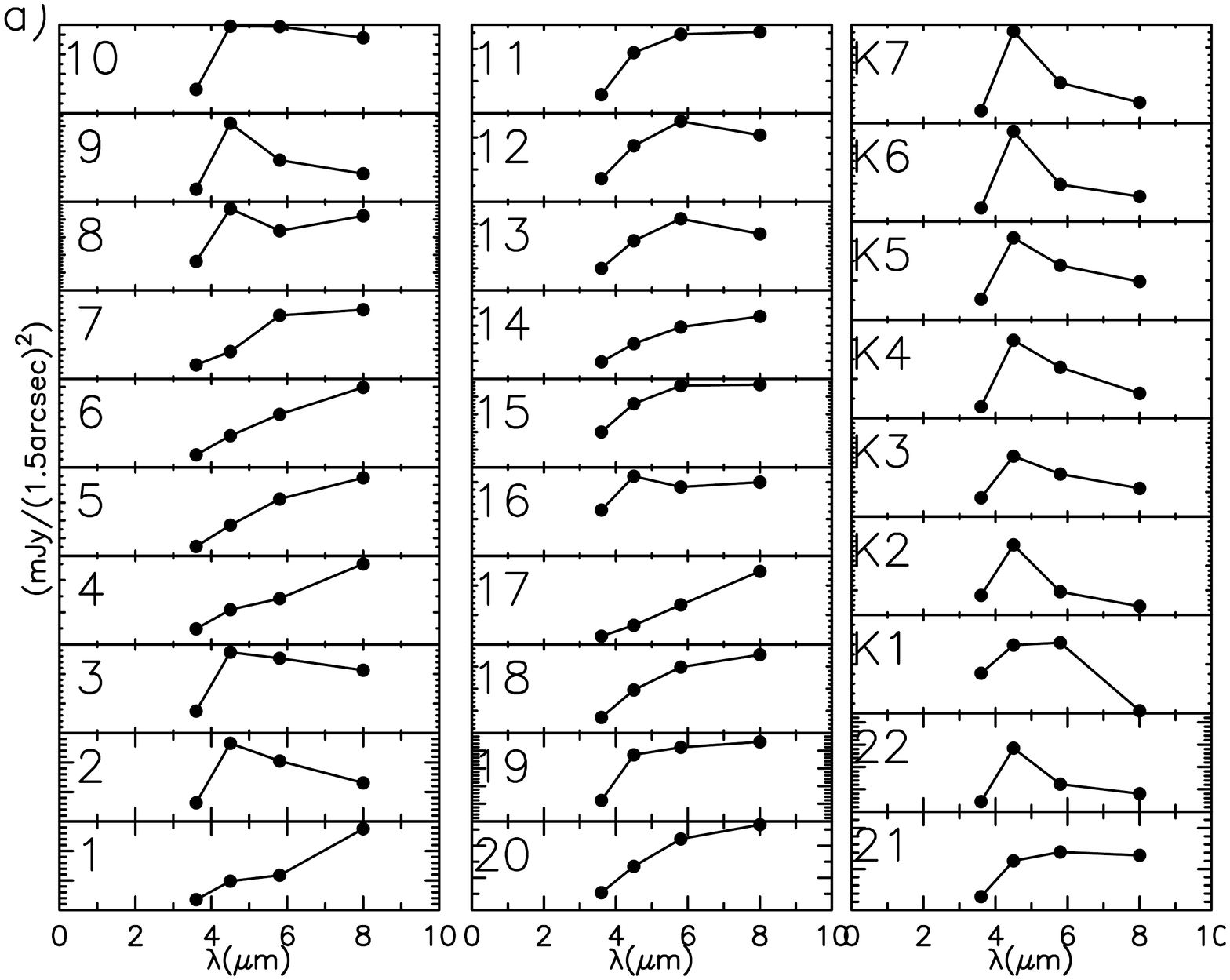}
\includegraphics  [ angle=-90,scale=0.5] {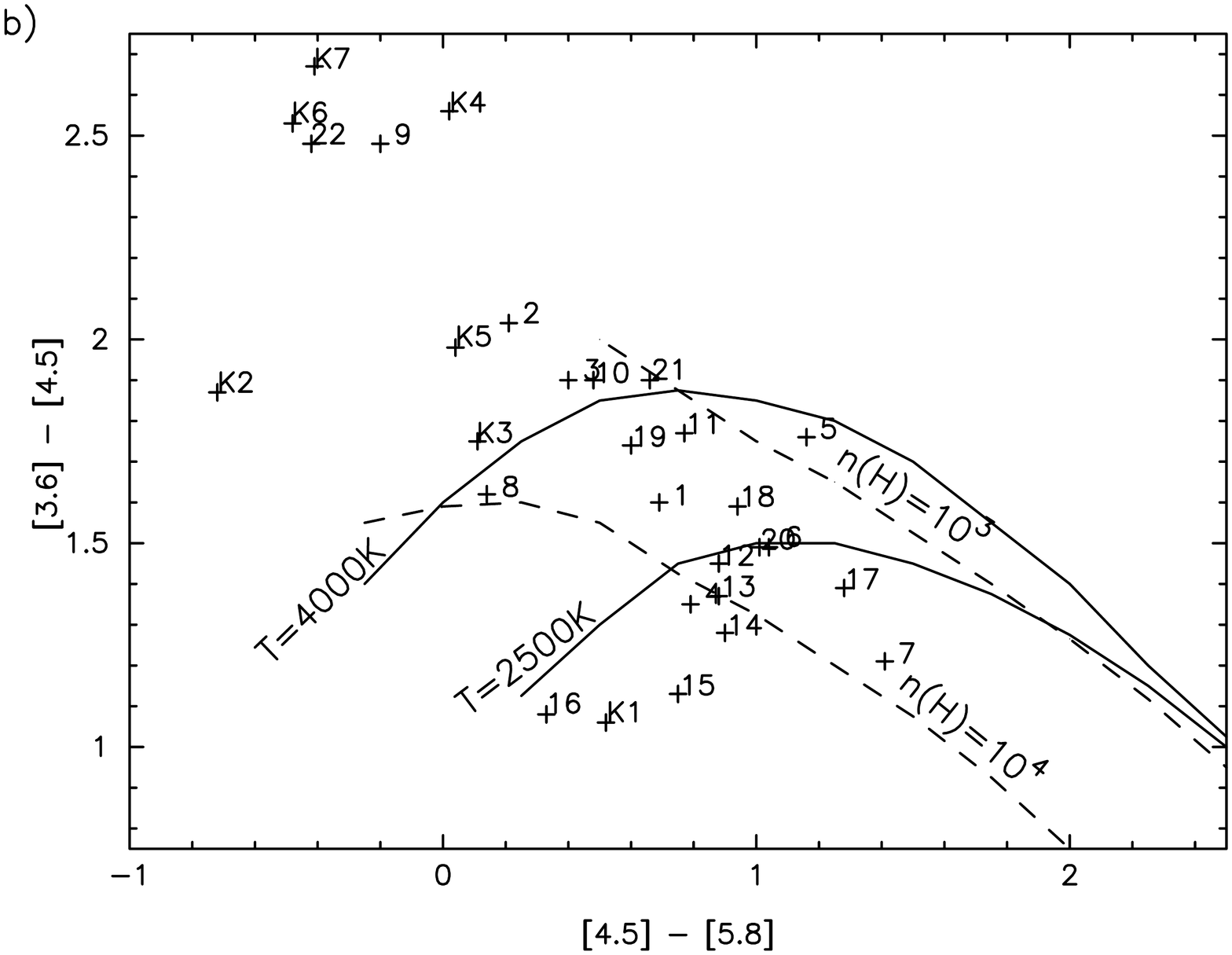}
\caption{ (a)  Flux density spectral plots of the  knots; the labels refer to the position of the knots  as marked on Figs. 5a \& 5b (see text). (b) IRAC color-color plot for the knots identified in Figs. 5a \& 5b.  The numbered labels  identify the knots. The constant temperature (solid lines) and density (broken lines),    reproduced from Ybarra \& Lada (2009), represent statistical equilibrium model calculations of  IRAC colors for \h2  gas in the shocks. \label{color-color}}
\end{figure}
\begin{figure}
\includegraphics  [scale=0.75] {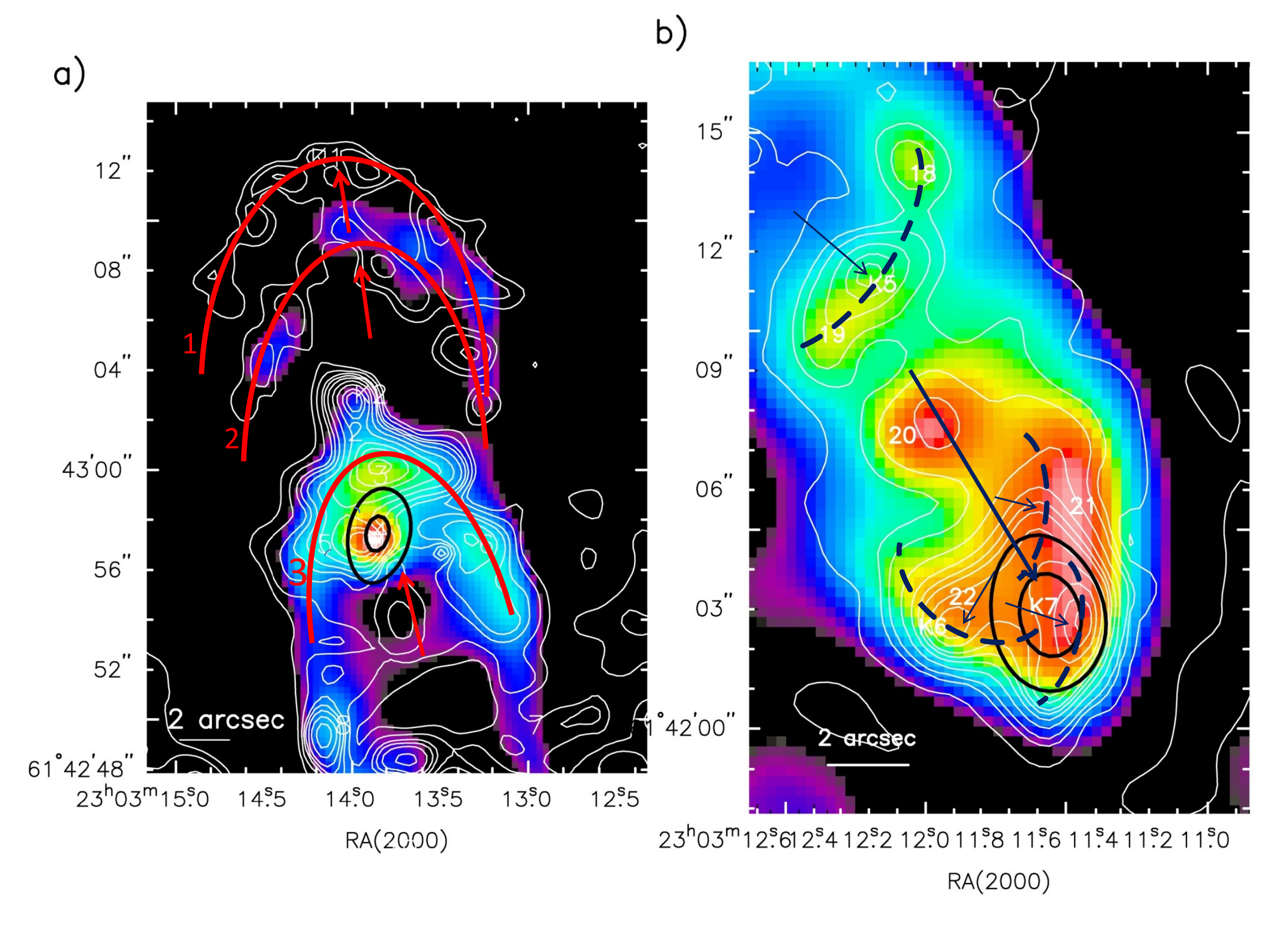}
\caption{The 4.5 \microns intensity contours (white) overlaid on 8 \microns emission (color image) showing the relative displacement between them. The   MIPS 24 \microns emission (tracing the atomic gas in the jet head/J- shocks) is shown by selected solid black contours (near peak and half power intensities). (a) Tip of NE lobe. Red arcs show a schematic of three bowshocks as inferred from 4.5 \microns emissions.  (b) Tip of SW lobe showing a complex system of oblique C-type shocks.   The arcs (dashed lines) represent leading edges of C-type shocks as inferred from the 4.5 \microns emissions.  The short arrows mark the ``putative''  projected oblique shock directions as ``inferred'' from the viewing geometry and the relative displacement between 8 \microns and 4.5 \microns intensities (see text). The long arrow  marks the direction of the atomic jet.    \label{geometry}}
\end{figure}
\begin{figure}
\includegraphics  [ scale=0.65, angle=-90]{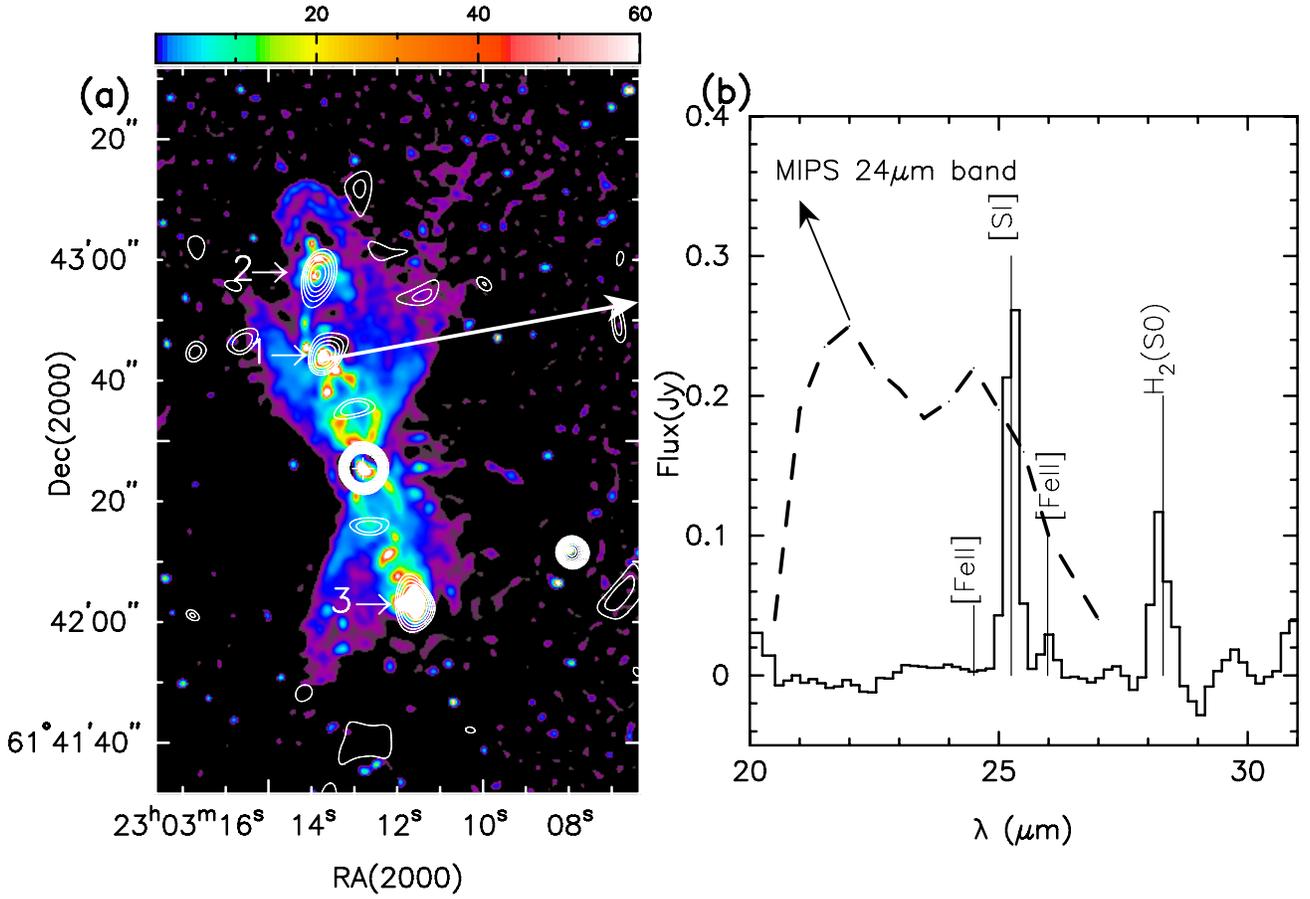} \caption{(a) 24 \microns emission (contours) overlaid on the 4.5 \microns image. The intensity contours are   at 4, 8, 16, 32, 64, 128, 256, 512, and 1024  MJy sr$^{-1}$. The contours near the center represent the protostar with a peak intensity of 47,597 MJy sr$^{-1}$. The long arrow marks the emission feature for which the IRS spectrum is shown. The labels 1 to 3 mark    the 24 \microns emission peaks along the jet. (b) IRS LL spectrum of the 24 \microns  NE emission peak (labeled \#1). The atomic/ionic  spectral features are marked.  The MIPS 24 \microns band response is also shown.  \label{24micron sources}}
\end{figure}
\begin{figure}
\includegraphics  [  scale=1.0] {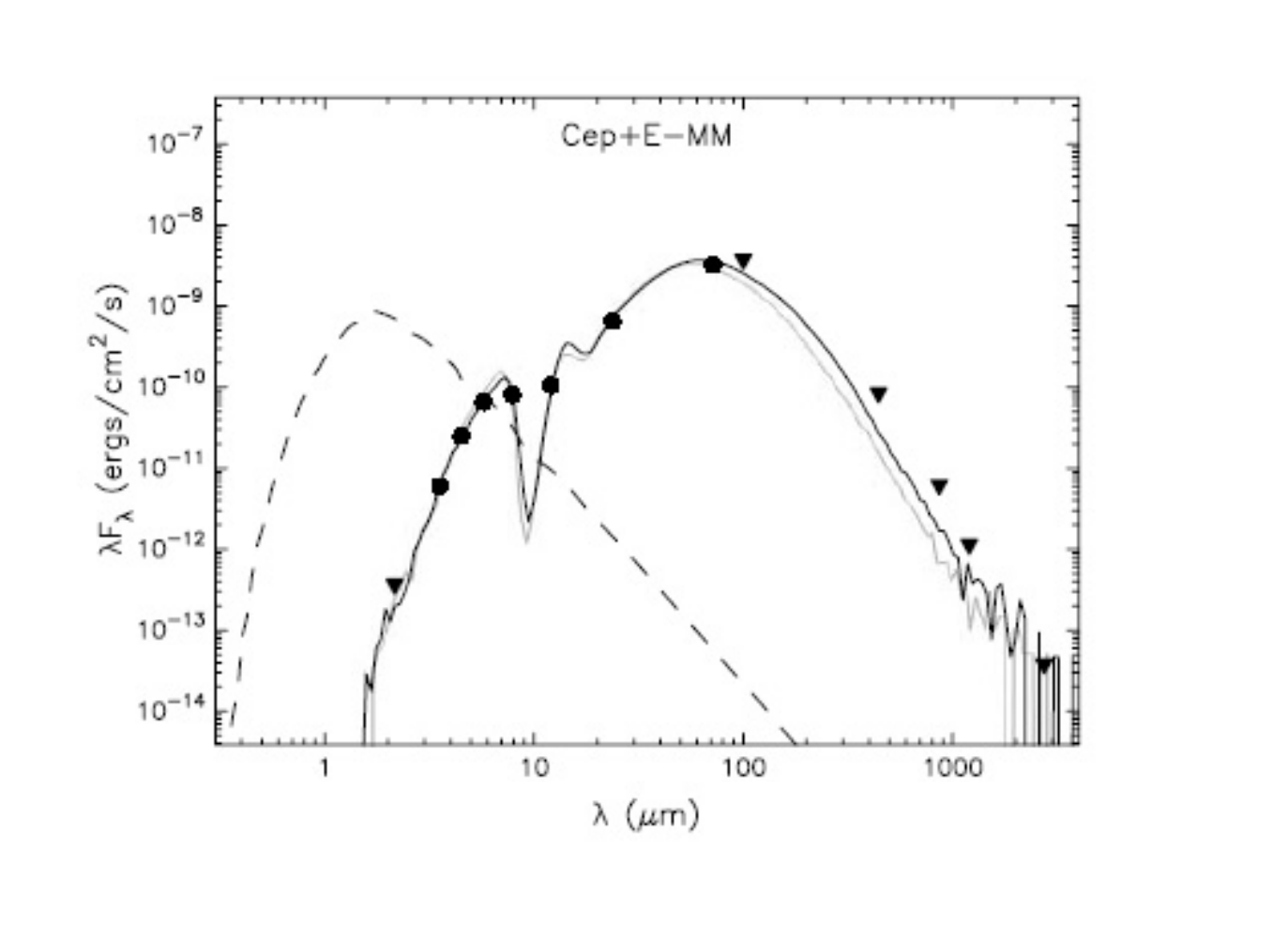}
\caption{ SED fitting for the Cep E protostar. The solid
black line shows the best fitting model. Grey lines show the models
that satisfy the criteria $\chi^{2} - {{\chi}^{2}}_{best} < 3$.  Black
dots are data points and triangles are upper limits. The dashed line
shows the stellar photosphere of the best fit model. \label{SED}}
\end{figure}
\clearpage
\begin{figure}
\includegraphics  [  scale=0.65, angle=-90]{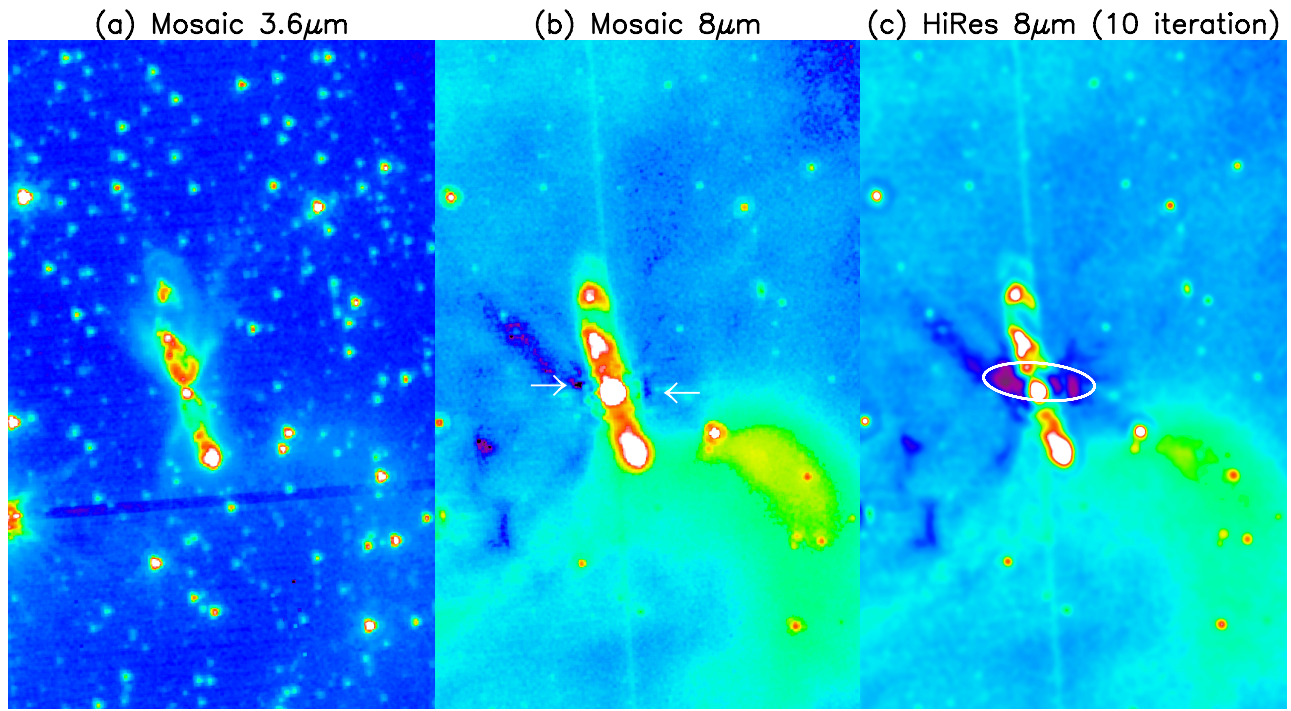} \caption{   Extinction at 8  \microns through a  dense envelope surrounding the protostar.   Mosaic images of a large region around the protostar showing the interstellar background at  3.6 \microns (panel a) and  at 8  \microns (panel b); the arrows (in panel b) mark extinction near the protostar.    (c) The 8  \microns HiRes deconvolved image after 10 iterations, processed without background subtraction (see text).   The ellipse indicates the flattened envelope region with high extinction at 8 \micron.
\label{darkcloud}}
\end{figure}

\end{document}